\documentclass[11pt]{article}
\usepackage[letterpaper,margin=1in]{geometry}
\usepackage{amsmath,amssymb,amsthm,mathtools}
\usepackage{enumitem}
\usepackage{microtype}
\usepackage{hyperref}
\hypersetup{colorlinks=true,linkcolor=blue,citecolor=blue,urlcolor=blue}
\usepackage{braket}
\usepackage{comment}
\usepackage{xcolor}
\usepackage{authblk}

\newtheorem{theorem}{Theorem}[section]

\newtheorem{lemma}[theorem]{Lemma}
\newtheorem{corollary}[theorem]{Corollary}

\newtheorem{remark}[theorem]{Remark}
\newtheorem{fact}[theorem]{Fact}
\newcounter{protocol}
\newtheorem*{maximizer-restated}
{Lemma~\ref{lem:packing-maximizer-exists}}

\newcommand{\R}{\mathbb R}
\newcommand{\C}{\mathbb C}
\newcommand{\bD}{\mathbb D}
\newcommand{\bT}{\mathbb T}
\newcommand{\cH}{\mathcal H}

\newcommand{\eps}{\varepsilon}

\newcommand{\iprod}[1]{\left\langle #1\right\rangle}
\newcommand{\norm}[1]{\left\| #1 \right\|}

\definecolor{gptnotecolor}{RGB}{128,0,128}

\title{Sensing frequencies at the discrete Grover-Heisenberg limit\\
using a single qubit}
\author{Boyang Chen\thanks{Department of Computer Science and Technology, Tsinghua University.  Email: \texttt{by-chen24@mails.tsinghua.edu.cn}} 
\quad 
Andrea Coladangelo\thanks{Paul G. Allen School of Computer Science and Engineering, University of Washington. Email: \texttt{coladan@cs.washington.edu}}
\quad 
Jerry Li\thanks{Paul G. Allen School of Computer Science and Engineering, University of Washington. Email: \texttt{jerryzli@cs.washington.edu}}
\quad 
Haocheng Zhang\thanks{Johns Hopkins University.}}
\date{}

\begin{document}
\maketitle

\begin{abstract}
We consider a fundamental quantum sensing task: detecting weak oscillating fields with unknown strength and frequency. Recent work by Allen et al.~\cite{allen2025quantum}
established a fundamental lower bound on the required sensing time,
which they named the ``Grover--Heisenberg limit'', and gave
a protocol attaining this limit. The protocol relies on auxiliary qubits, which are used to run Grover's algorithm as a subroutine, leaving open the question of whether a \emph{single} sensing qubit can achieve the same scaling. 

We show that a single qubit suffices for a discrete version of this
problem in which the frequency
belongs to an \emph{arbitrary}, but known, finite set of candidate frequencies. Our protocol's sensing time grows only as the square root of the number of
candidate frequencies. We prove that this square-root dependence, which we refer to as the ``discrete Grover-Heisenberg limit'', is optimal in the worst case.
\end{abstract}

\medskip
\medskip
\medskip
\medskip
\begin{center}
\setlength{\fboxsep}{8pt}
\fbox{%
\parbox{0.90\textwidth}{%
     \small
     \textbf{AI usage statement.} The authors came up with preliminary ideas for how one might obtain a sensing protocol using a single qubit. These ideas were fleshed out and expanded upon with help from GPT 5.6. The authors then came up with the idea of considering the setting of discrete frequencies, and reduced the problem to the ``simultaneous correlation’’ problem described in the paper. This crucial problem was resolved entirely by GPT 6 (Theorem~\ref{thm:main}), along with its upgrades in Section~\ref{sec:sign-packing} and \ref{sec:random-delay}. The authors then additionally conjectured a matching “discrete Grover-Heisenberg limit”. This conjecture turned out to be correct, and was also proved by GPT~6 (Theorem~\ref{thm:discrete-sensing-lower-bound}). These proofs were substantially revised by the authors for clarity and completeness. The authors also devoted substantial effort to the writing and exposition of the paper, with assistance from GPT~6. 
   }
}
\end{center}

\newpage
\tableofcontents

\section{Introduction}
Quantum sensing uses quantum systems to detect weak signals and
measure physical quantities with a precision that can go beyond
conventional limits~\cite{degen2017quantum,giovannetti2004quantum}.
It has emerged as a promising application of quantum devices,
with uses ranging from magnetic-field measurements and
nanoscale spectroscopy~\cite{degen2017quantum} to biomedical sensing~\cite{aslam2023quantum}, atomic
clocks~\cite{aeppli2024clock} and gravitational-wave
detection~\cite{ganapathy2023broadband}.

A fundamental problem in quantum sensing is \emph{broadband AC sensing}: detecting weak oscillating fields with unknown strength, phase, and frequency. Informally, the task is to decide
whether a signal is present, given a lower bound on its strength and a range of possible frequencies.
When the frequency is known, control pulses timed to its period can make the signal accumulate coherently. When it is unknown, a conventional approach is to scan the possible frequencies by varying the period of the control
pulses~\cite{carr1954effects,meiboom1958modified}, which becomes costly as the frequency range grows.

Recently, Allen et al.~\cite{allen2025quantum} introduced
\emph{quantum computing enhanced} sensing, showing how Grover's
search algorithm~\cite{grover1996fast} can speedup this task.
Their idea is to convert evolution under the unknown signal
into an approximate reflection marking the frequency interval that
contains the signal. Alternating this operation with a reflection
about an initial superposition implements a Grover search over the
frequency intervals, giving a quadratic improvement in the
dependence of the sensing time on the width of
the possible frequency range.
They also establish a fundamental lower bound
on the sensing time, matching what is achieved by their protocol, which they name the \emph{Grover--Heisenberg
limit}. Recent works by Xu et al~\cite{xu2026experimental} and Ouellet et al.~\cite{ouellet2026experimental} have performed experimental demonstrations of this approach using superconducting qubits coupled to microwave cavities, showing an improvement over conventional approaches.

Now, unlike conventional approaches, the protocol of Allen et al.\ uses auxiliary qubits to store the
frequency intervals in superposition and run Grover's algorithm. Thus, it requires coherent control of a quantum register in addition to the sensing qubit itself. The authors leave open the question of whether the same
speedup is possible without auxiliary qubits, i.e.\ with only one sensing qubit~\cite{allen2025quantum}. In this work, we focus on this question:
\begin{center}
\begin{minipage}{0.88\linewidth}
\centering
\emph{Can the Grover--Heisenberg limit be achieved\\
with a single sensing qubit and no auxiliary qubits?}
\end{minipage}
\end{center}

\subsection{Our results}
We show that a single sensing qubit suffices for a
\emph{discrete} version of broadband AC sensing, in which the unknown frequency
belongs to an arbitrary, but known, finite set of candidate frequencies. We present a protocol where the sensing time grows as the square root of the number of candidate frequencies. The protocol alternates instantaneous single-qubit pulses with evolution under the signal Hamiltonian. 

More precisely, our task is to distinguish the zero Hamiltonian
from a signal Hamiltonian of the form
\[
 H_{\mathrm{sig}}(t)=B\cos(\omega t+\phi)Z,
\]
where $Z$ is the Pauli $Z$ operator acting on the sensing qubit,
$B>0$ is the signal strength, $\omega$ is its frequency, and
$\phi$ is its phase. We are given as input a set
$W=\{\omega_1,\ldots,\omega_N\}\subset(0,\infty)$ containing
$\omega$ and a positive lower bound $B_{min}\leq B$, as well as access to evolution under $H_{\mathrm{sig}}(t)$, but the
actual strength, frequency, and phase are otherwise unknown.\footnote{For the unfamiliar reader, this detection problem might appear almost trivial at first sight: why not initialize the sensing qubit in the state $\ket{+}$, evolve it under the signal Hamiltonian, and check if the final state is still $\ket{+}$ or has changed? The difficulty is that, in order for the initial state to change by a constant amount, one needs to evolve for a time $\sim1/B$, where $B$ is the strength of the signal. However, in the regime where $\omega >> B$, the integral of the cosine function over an interval of size $1/B$ is almost zero due to the cosine's oscillation.}
\begin{theorem}[Informal]
\label{thm:intro-upper}
For every $B>0$, $N \in \mathbb{N}$, there is a protocol satisfying the following: given as input  a set of candidate frequencies $W = \{\omega_1, \ldots, \omega_N\}$, and a lower bound $B_{min} \leq B$, the protocol uses one sensing qubit (and no auxiliary qubits) and detects the signal with error probability at most
$1/3$, for every candidate frequency $\omega \in W$ and every phase $\phi$, in total time
$$
 O\!\left(\frac{\sqrt N}{B_{min}}+\frac1{\omega_{\min}}\right),
$$
where $\omega_{\min}:=\min_i\omega_i$.\footnote{The regime where the sensing problem is difficult is typically the ``high-frequency'' regime, where  $\omega >> B$. In this regime, the additive $1/\omega_{min}$ term in the total time is subsumed by the first term. Other methods are available in the ``low-frequency'' regime, where the Hamiltonian becomes approximately time-independent.}
The protocol uses finitely many Pauli $Y$ gates, alternating with evolution under the signal Hamiltonian.
\end{theorem}
We emphasize that, perhaps surprisingly, no assumption is needed on the set of candidate frequencies, which can be arbitrary. At the heart of our protocol is a result about how a single
bounded function can correlate with many oscillations at once. Given $N$ frequencies $\omega_1,\ldots,\omega_N$, the problem is to find a function $u$ for which the correlations
\[
 \int_0^{2\pi}u(t)\cos(\omega_i t)\,dt,
 \qquad i=1,\ldots,N,
\]
all have relatively large magnitude, at least $\Omega(1/\sqrt{N})$, while keeping $|u(t)|\leq1$ at every point. We refer to this as the ``simultaneous correlation'' problem. Our main technical result is that such a function $u$ always exists, for arbitrary sets of candidate frequencies. This simultaneous correlation problem is related to classical
questions in harmonic analysis and discrepancy theory, but asks
for a different guarantee. Discrepancy theory typically seeks sign choices that make many sums small. Here, we seek a function
that remains bounded at \emph{every} point while having large correlations with all the prescribed cosines. Standard discrepancy results do not appear
to apply directly, and our argument instead uses tools from complex analysis. Both the result and the techniques may be of
independent interest.

We also prove that the square-root dependence on the number of
candidate frequencies is necessary in the worst case. We refer to
the corresponding lower bound as the \emph{discrete Grover--Heisenberg limit}.
\begin{theorem}[Informal]
\label{thm:intro-lower}
For every $B>0$ and every positive integer $N$, there
exists a set $W$ of $N$ candidate frequencies such that the following holds: 
any protocol solving the corresponding sensing problem with error probability at most $1/3$ on every frequency $\omega \in W$ requires time
\[
 \Omega\!\left(\frac{\sqrt N}{B}\right),
\]
This holds even with
arbitrary controls and arbitrarily many auxiliary qubits
in addition to the sensing qubit, and even when $B$ and the phase $\phi$ are known.
\end{theorem}

The proof of the latter can be seen as a discretization of the lower bound in \cite{allen2025quantum}, which establishes the Grover-Heisenberg limit for sensing frequencies from a continuous window.

The reader might wonder whether the original broadband AC sensing problem, with a continuous window of candidate frequencies, can also be solved at the Grover-Heisenberg limit with a single sensing qubit (where recall that \cite{allen2025quantum} achieves this limit, but using auxiliary qubits). Preliminary investigations by the authors suggest that this is not possible, and that a conventional sensing lower bound holds for a broad class of protocols that use only constantly many qubits. However, we do not present any formal proof here, and we leave this exploration for future work.

\paragraph{Practical considerations for implementating our protocol.}
Our protocol assumes that a finite set of candidate frequencies is known. In reality, these need not be known exactly, but they should be known to precision $cB_{min}/N$, for a sufficiently small universal constant $c$. 
Such an approximation error would preserve our sensing time scaling up to constant factors.
This kind of discrete frequency assumption is not suitable for every sensing setting, but
can be motivated by prior knowledge of the signal source. For example, characterizing the nuclear spins near a nitrogen-vacancy (NV) center involves identifying which nearby diamond lattice sites contain $^{13}\mathrm C$ nuclei. Their positions determine their ``hyperfine couplings'' to the NV center and hence their resonance frequencies~\cite{dreau2012high,zopes2018three}.
If we consider a finite set of nearby sites and can predict their resonance frequencies accurately, we obtain a finite list of candidates.

The second requirement of our protocol concerns control timing. Let
$\omega_{\max}:=\max_i\omega_i$. In the regime
$\omega_{\min}\geq B_{min}$, our construction uses $O(\omega_{\max}N/B_{min})$ pulses in total. This is the same worst-case order
as a conventional scan that interrogates each candidate frequency for
time $\Theta(1/B_{min})$ using pulses timed to its period. However, note that our protocol applies its pulses over a shorter
$\Theta(\sqrt N/B_{min})$ sensing time compared with $\Theta(N/B_{min})$
for a conventional scan. This means that the average pulse-rate in our protocol is
$O(\omega_{\max}\sqrt N)$ rather than $O(\omega_{\max})$, i.e.\
a factor of $\sqrt N$ higher. Depending on the setting, this might present a challenge that needs to be overcome for practical implementations. 

\paragraph{Recent related work.} 
We mention here a recent related work of Zhang et al.~\cite{zhang2026moving}, which considers a different
setting with $N$ possible discrete alternatives. There, a field of known strength $B$ occupies one of $N$ spatial locations. They show that, by allowing a sensor to occupy a coherent superposition of locations, one can achieve sensing time
$\Theta(\sqrt N/B)$ in their setting, with a matching lower bound. Their
protocol uses a single sensor, but its $N$-dimensional position space
serves as a quantum register on which Grover search is implemented. In our setting, the $N$ alternatives are frequencies, and the entire
quantum system used by the protocol is a single-qubit sensor, without any additional position or frequency register. 

We mention another related work work by
Polloreno and Smith~\cite{polloreno2026simple}, building on \cite{polloreno2023opportunities}, which also studies broadband AC sensing over a continuous frequency interval. They give an alternative derivation of the Grover--Heisenberg
limit of Allen et al.~\cite{allen2025quantum} using bounds on the integrated quantum Fisher information,
that is, the quantum Fisher information integrated over the
possible signal frequencies. They also propose an analog
protocol approaching this limit. This protocol, like that of Allen et al., requires an auxiliary quantum system that grows with the frequency bandwidth. Our result instead handles a finite set of candidate frequencies, but without needing any auxiliary system.

\subsection*{Acknowledgements}
A.C.\ is supported in part by the NSF under CAREER Award No.~2541085, and by the Google Research Scholar program. J.L.\ is supported in part by the NSF AI Institute for Foundations of Machine Learning (IFML). H.Z.\ was supported by the Quantum@UW REU program, funded by NSF Award No.~CCF-2446908.

\section{Technical overview}
\label{sec:technical-overview}
Our task is to distinguish the zero Hamiltonian from a signal $H_{sig}(t) = B\cos(\omega t+\phi)Z$, given a set of $N$ candidate frequencies containing $\omega$ and a
lower bound $B_{min}\leq B$. We are also given access to evolution under the signal Hamiltonian, as well as the ability to apply ``control'' unitaries on the sensing qubit. The oscillation of the signal is what makes the problem difficult: in order for the initial state to change by a constant amount when evolving under this Hamiltonian, one needs to evolve for a time $\sim1/B$, where $B$ is the strength of the signal; however, in the regime where $\omega >> B$, the oscillation of the cosine function makes the integral over an interval of size $1/B$ almost zero. So, simply letting the qubit evolve for some time does not produce a detectable change. 

Now, when the frequency $\omega$ is known, one can apply an appropriate sequence of control ``pulses'' (i.e.\ unitaries) with period $1/\omega$ that flip the effect of the signal whenever its sign is negative. This lets the contributions accumulate constructively, producing a change on the order of $BT$ in the initial state (say, $\ket{+}$), where $T$ is the sensing time, i.e.\ the time spent evolving. Thus, the natural sensing time scale to achieve a constant change is $T \sim 1/B_{min}$. Of course, the difficulty is that $\omega$ is unknown, except for the fact that it lies in the set of $N$ candidate frequencies. The conventional approach~\cite{carr1954effects,meiboom1958modified} is to essentially do a brute-force search over all the candidate frequencies, devoting time $\sim 1/B_{min}$ to each, for a total sensing time of $O(N/B_{min})$.

Our goal will be to obtain a constant change without testing the candidate frequencies one at a time. We show how to do this with a \emph{single} sensing qubit in time $O(\sqrt N/B_{min})$. Crucially, our protocol is not based on Grover search, as this would require using auxiliary qubits. For simplicity, we will focus first on the case where the phase $\phi =0$, and we will describe at the end how to deal with the case of an unknown phase $\phi$.

\paragraph{Expressing what the control must accomplish.}
In general, we can describe our control action in the sensing protocol by a Hamiltonian $H_c(t)$. Then, the total Hamiltonian is $H_{tot}(t) = H_{sig}(t) + H_c(t)$. Denote by $U_c(t)$ the unitary evolution induced by $H_c(t)$. Denote by $\ket{\psi(t)}$ the state of the qubit at time $t$.

It is helpful to study what happens in the ``control-following basis'' (also known as the interaction picture) i.e.\ in the basis $\ket{0_c(t)}:=U_c(t)\ket{0}$, $\ket{1_c(t)}:=U_c(t)\ket{1}$. These basis states evolve exactly as if the signal Hamiltonian were absent. We can then write
\[
|\psi(t)\rangle
=
a(t)|0_c(t)\rangle+b(t)|1_c(t)\rangle.
\]
for some time-dependent coefficients, and define the coefficient vector  $|c(t)\rangle
=
a(t)|0\rangle+b(t)|1\rangle$.
This satisfies the relation
$\ket{\psi(t)}=U_c(t)\ket{c(t}\rangle$. Moreover, suppose we initialize the sensing qubit in the state $\ket{0}$. Then, we have $\ket{\psi(0)} = \ket{c(0)} = \ket{0}$. Now, when the signal is zero, the only evolution is due to the control Hamiltonian, and thus $\ket{c(t)}$ stays constant and equal to the starting state $\ket{0}$. On the other hand, when the signal is present, it is straightforward to show that the coefficient vector satisfies the following ``interaction picture'' Schrodinger equation, which we derive in Section~\ref{sec:analysis_one}:
\begin{equation}
\label{eq:interaction}
 i\frac{d}{dt}\ket{c(t)}
 =B\cos(\omega t)M(t)\ket{c(t)},
 \qquad M(t)=U_c(t)^\dagger ZU_c(t) \,.
\end{equation}
The actual equation from Section~\ref{sec:analysis_one} has $\cos\bigl(\omega(t+\tau)\bigr)$ for an additional parameter $\tau$, where $\tau$ is a ``wait time'' introduced in our protocol before we start the signal evolution (in order to handle the unknown phase $\phi$). We ignore this in this overview for simplicity.

So, the goal is to design an (implementable) $M(t)$ that, after time $T\sim \sqrt{N}/B_{min}$, results in $c(T)$ having \emph{constant} weight on $\ket{1}$. If this can be achieved, then undoing the known control unitary $U_c(T)$ and making a standard basis measurement will result in a constant probability of detecting the signal.

We will consider controls of the form $M(t)=f(t)X$, where $f(t)$ is some bounded measurable function. While much more general controls are possible, we will show that this class of controls already suffices for our purposes. The first question is: what kinds of function $f(t)$ lead to an implementable $M(t)$? Notice that, since conjugation preserves eigenvalues, $M(t)$ has the same eigenvalues as $Z$, i.e.\ $\pm1$. So, $f(t)$ must take values in $\{-1,1\}$. This is quite a restrictive condition! It says that, at best, what we get to choose is when to switch between $X$ and $-X$. Note further, by a direct calculation, that we can achieve $U_c(t)^\dagger ZU_c(t) = \pm X$ by choosing $U_c(t) = e^{\pm i\pi Y/4}$. So, indeed, any piecewise-constant function $f(t)$ taking values in $\{\pm1\}$ is implementable via a sequence of control unitaries that flip $e^{i\pi Y/4}$ to $e^{-i\pi Y/4}$ and viceversa, at each time where $f(t)$ changes sign. One can readily verify that such control unitaries are actually just $Y$, up to a global phase! 

Now, importantly, our choice of $M(t)$ is such that $M(t)$ and $M(t')$ commute for all $t,t'$. So, the solution to Equation \eqref{eq:interaction} has the simple form\footnote{Without commutation, the solution would be a more general Dyson expansion.}
$$
 \ket{c(T)}=e^{-i A_\omega(T)X}\ket0
 =\cos \left(A_\omega(T) \right)\ket0-i\sin \left( A_\omega(T)\right)\ket1 \,,
$$
where 
$$
 A_\omega(T) := B\int_0^T f(t)\cos(\omega t)\,dt \,.  $$
Thus, $\ket{c(T)}$ has constant weight on $\ket{1}$ when $\sin^2(A_\omega(T))$ is constant, or, equivalently, when $A_\omega(T)$ itself is constant (technically, $A_\omega(T)$ should also be bounded away from multiples of $\pi$ by a constant -- this is a small detail that is handled by the random wait time $\tau$ mentioned earlier, which we are ignoring in this overview). Crucially, we wish to find an $f(t)$ such that the latter holds \emph{for all} $\omega$ in the set of candidate frequencies $\{w_1, \ldots, w_N\}$, and we want this to happen at $T \sim \sqrt{N}/B_{min}$.

\paragraph{A ``simultaneous correlation'' problem.}
It is convenient to make the change of variables $s=2\pi t/T$, and to set
$w_i=\omega_iT/(2\pi)$, and $u(s)=f(Ts/(2\pi))$. Then,
$$
 A_{\omega_i}(T)
 =\frac{BT}{2\pi}\int_0^{2\pi}u(s)\cos(w_i s)\,ds.
$$
This rewriting makes the required scale explicit: at
$T=\Theta(\sqrt N/B_{min})$, an integral of magnitude
$\Omega(1/\sqrt N)$ gives an $A_{\omega_i}(T)$ with at least constant magnitude. Thus, the problem we wish to solve has now become:
given a set of arbitrary candidate frequencies $\{w_1, \ldots, w_N\}$, find a measurable piecewise-constant function $u: [0,2\pi] \rightarrow \{\pm1\}$ such that
\[
 \left|\int_0^{2\pi}u(s)\cos(w_i s)\,ds\right|
 = \Omega\left(\frac{1}{\sqrt{N}}\right) \qquad\textnormal{for all }1\leq i\leq N.
\]
Now, while we do not solve this exact problem, it turns out that it suffices for us to solve the following \emph{relaxation} of the problem, where the function $u$ is only required to be measurable and bounded (but not necessarily piecewise-constant with $\pm1$ values):
\begin{center}
\begingroup
\setlength{\fboxsep}{10pt}
\setlength{\fboxrule}{0.5pt}
\fcolorbox{blue!35!black}{blue!2}{%
\begin{minipage}{\dimexpr\linewidth-2\fboxsep-2\fboxrule\relax}
Given a set of arbitrary candidate frequencies $\{w_1, \ldots, w_N\}$, find a measurable function $u: [0,2\pi] \rightarrow \C$, with $|u(s)|\leq1$ for every $s\in[0,2\pi]$, such that
\begin{equation}
 \left|\int_0^{2\pi}u(s)\cos(w_i s)\,ds\right|
 = \Omega\left(\frac{1}{\sqrt{N}}\right) \qquad\textnormal{for all }1\leq i\leq N.
 \label{eq:overview-correlation}
\end{equation}
\end{minipage}%
}
\endgroup
\end{center}
Solving the latter problem is the technical crux of our result. Later, we describe how to convert such
a function $u$ into a \emph{pair} of piecewise-constant functions $f_0, f_1: [0,2\pi] \rightarrow \{\pm1\}$ such that, for each $i$, at least one of the two satisfies the desired correlation bound. The existence of such a pair is good enough for our purposes as we can simply choose one of the two functions uniformly at random in our protocol, at the cost of only a constant factor loss in detection probability. 
So, we shall focus on the boxed problem.

First, we start by getting some sense for the scale of the desired correlations. Imagine for a moment that the cosines were orthogonal, in the sense that $\int_0^{2\pi}\cos(w_i s)\cos(w_j s)\,ds=0$ whenever $i\ne j$. This happens for example when the $w_i$ are distinct
positive integers (in our problem they can be arbitrary real frequencies, but let us assume orthogonality for now). A natural first attempt is to define $u$ to be a uniform linear combination of each cosine, i.e.\ $u(s)=\frac{1}{\sqrt N}\sum_{j=1}^N\cos(w_j s)$. Orthogonality makes the cross terms vanish, so
$$
 \int_0^{2\pi}u(s)\cos(w_i s)\,ds=\frac{\pi}{\sqrt N} \qquad \textnormal{for all } i \,.
$$
This looks like the desired solution, except that it is not bounded: $u(0)=\sqrt N$ since all the
cosines line up constructively. Replacing
the weights by $1/N$ makes the function bounded by one everywhere,
but then each correlation is only $\pi/N$, which would lead to a sensing time of order $N/B_{min}$.
 
In our problem, the frequencies are arbitrary, so orthogonality does not hold in general and we cannot even rely on the cross terms vanishing. This problem is reminiscent of problems in discrepancy theory~\cite{matousek1999geometric,chazelle2000discrepancy}, where one wishes to choose
signs in such a way that many signed sums are small simultaneously. In the orthogonal
example, one could try replacing the positive coefficients by signs:
if $\sum_{j=1}^N\varepsilon_j\cos(w_j s)$, with
$\varepsilon_j\in\{\pm1\}$, were bounded by $O(\sqrt N)$ at every
time, normalizing it would give a bounded function with every
correlation of magnitude $\Omega(1/\sqrt N)$. For example, Spencer's theorem~\cite{spencer1985six} says that we can indeed find a choice of signs so that $O(N)$ sums are all simultaneously at most $O(\sqrt N)$. This implies the following. For any choice of
$O(N)$ times $t_1, \ldots, t_{O(N)}$ in $[0,2\pi]$, there is a choice of signs such that all $O(N)$ sums (where the the $j$-th sum has the cosine values at $t_j$ as coefficients) are controlled by $O(\sqrt N)$. Unfortunately though, we want to control the sums at \emph{all} times in the interval $[0,2\pi]$, and Spencer's theorem does not give any guarantee that there are no large peaks between the chosen times. 
Moreover, we are again assuming orthogonality, which may not hold in general. More broadly, it does not seem that known results from discrepancy theory are able to provide the guarantees we are looking for.

\paragraph{Solving the simultaneous correlation problem.}
We take a different approach to finding a single bounded function whose correlation with \emph{each} cosine is large. We approach this by an optimization
over all measurable functions $u:[0,2\pi]\to\C$ with $|u(s)|\leq1$. Crucially, the variable we optimize is the function $u$ itself. We will choose an objective functional and prove that the function maximizing it has the required
correlations. This is the strategy behind Theorem~\ref{thm:main}.

The first choice we have to make is how to combine the $N$ correlation magnitudes
into one single objective. For example, their sum does not adequately capture our goal:
several large correlations could hide a very small one. A natural starting point is instead their \emph{product}, which vanishes whenever a single one of them vanishes. We further take its logarithm to express it as a sum, which will make the later comparisons easier. So, we will consider optimizing the functional
$$ J(u) =
 \log\!\left(\prod_{i=1}^N
   \left|\int_0^{2\pi}u(s)\cos(w_i s)\,ds\right|\right)
 =\sum_{i=1}^N\log\left|
   \int_0^{2\pi}u(s)\cos(w_i s)\,ds\right|.
$$

We can now make the proposed argument more concrete. Suppose a
maximizing function $u$ has a correlation that is smaller than the bound
we want. To obtain a contradiction, we would need to find another
function, still bounded by one, with a larger objective. The natural
attempt is to modify $u$ to increase this small correlation. However, the
same $u$ appears in every integral, so this modification also affects
the other correlations. Thus, we must compare the gain in the selected
correlation with any losses in the others. The logarithms make this
comparison convenient: doubling the selected correlation gives a
gain of $\log 2$, regardless of how small it was initially.

The difficulty with such a modification is again to preserve the boundedness of $u$ (recall that we need $|u(s)|\leq 1$ for all $s$). If $u$ is already close to magnitude
one over much of the interval, adding even a small function to it may
violate this bound. To obtain a maximizer with room for modification, the idea is to adjust the objective by adding a penalty or ``regularizing'' term. We use
$\log(1-|u(s)|^2)$ for such a term, which is zero at $u(s)=0$ and becomes increasingly
negative as $|u(s)|$ approaches one. Its integral penalizes functions
that are close to magnitude one over much of the interval. Our
final objective is then
\[
 \tilde{J}(u) = \sum_{i=1}^N\log\left|
       \int_0^{2\pi}u(s)\cos(w_i s)\,ds\right|
       +CN\int_0^{2\pi}\log(1-|u(s)|^2)\,ds,
\]
where $C$ is a sufficiently large universal constant.

To see why the penalty helps, consider shrinking a maximizer $u$
slightly, by multiplying it by a factor $\beta<1$. Every correlation shrinks by the same factor, so each of
the $N$ logarithmic terms loses the same amount. At the same time,
the penalty becomes less negative. We give it weight proportional
to $N$ so that it can balance the loss from all $N$ correlations.
If $u$ had large magnitude over too much of the interval, the
improvement from decreasing the penalty term would outweigh that loss, contradicting
maximality. Lemma~\ref{lem:packing-energy-bound} makes this comparison precise: a maximizer $u$ must satisfy
\begin{equation}
\label{eq:bound}
\int_0^{2\pi}|u(s)|^2\,ds\leq\frac{1}{2C}.
\end{equation}
In particular, $|u(s)|\leq1/2$ except on a set of total length at
most $2/C$. This gives us room to modify $u$ on most of the interval.

We now need to choose the modification itself. Fix a candidate $i$
whose correlation\\ $\left|\int_0^{2\pi}u(s)\cos(w_i s)\,ds\right|$ at the maximizer $u$ is supposedly too small, and write
$\sigma_i(s)=\operatorname{sgn}(\cos(w_i s))$. This sign function is a natural way to target the
$i$-th integral: adding $\sigma_i(s)$ to $u(s)$, for some complex number $z$, changes that integral
by $\int_0^{2\pi}|\cos(w_i s)|\,ds$. Here, we give ourselves an extra degree of freedom, and we consider adding $z\sigma_i(s)$ for some complex number $z$, which changes the integral
by $z\int_0^{2\pi}|\cos(w_i s)|\,ds$. The presence of $z$ will be important, as it will allow us to vary both the size and the direction of the change. 

Now, the problem with this simple addition is that it may still push $u(s)$
outside of the unit disc. To keep the modified values inside the disc, we use a standard
construction from complex analysis. For $|a|<1$, we consider the M\"obius
transformation $\varphi_a(w)=(w-a)/(1-\overline a w)$, which maps the
unit disc bijectively onto itself and sends $a$ to $0$. 
Then, fix a time $s$. We want a modification that stays inside the disc
and still follows the direction $z\sigma_i(s)$ of our original
additive attempt. The inverse map $\varphi_a^{-1}$ is useful for both purposes:
it maps the disc to itself, and one can check that for small $v$ it satisfies
\[
 \varphi_a^{-1}(v)=\frac{a+v}{1+\overline a v}
 =a+(1-|a|^2)v+O(|v|^2).
\]
The factor $1-|a|^2$ is positive, so, to first order, the change from $a$
is in the same direction as $v$, just with a smaller magnitude
near the boundary. Motivated by this, we take $a=u(s)$ and $v=z\sigma_i(s)$, and define our modified function to be
$$
 u_z(s)=\varphi_{u(s)}^{-1}(z\sigma_i(s))
       =\frac{u(s)+z\sigma_i(s)}
              {1+z\overline{u(s)}\sigma_i(s)},
 \qquad |z|<1.
$$
Each $u_z$ satisfies $|u_z(s)|\leq1$ everywhere, and setting $z=0$
recovers the original function: $u_0=u$.

How does this modification affect the $i$-th correlation? A direct calculation (which we postpone to Section~\ref{sec:analysis_packing}) gives
\begin{align*}
 \int_0^{2\pi}u_z(s)\cos(w_i s)\,ds
 &=\int_0^{2\pi}u(s)\cos(w_i s)\,ds\\
 &\quad+z\int_0^{2\pi}(1-|u(s)|^2)|\cos(w_i s)|\,ds
       +O(|z|^2).
\end{align*}
The key point here is that the earlier bound on $\int|u|^2$ from \eqref{eq:bound} keeps the
coefficient of $z$ above a positive constant. Thus, we still achieve a change of order
$|z|$ in the target integral. 

How should we choose $z$? First, fix a small magnitude $r$ and leave its
direction free. The linear change then has magnitude proportional
to $r$, while the error is only $O(r^2)$. Take $r$ small enough so that
the error term has magnitude at most, say, one quarter of the linear term's
magnitude. \emph{If} the original correlation is also at most one quarter
of that magnitude, the original integral and the error together can
cancel at most half of the linear term, whatever the direction of
$z$. The new correlation then
has at least twice its original magnitude for every $|z|=r$.
We therefore gain at least $\log 2$ in the $i$-th correlation, regardless of the direction of $z$.

There are two possible costs though: the other correlations (for $j\neq i$) may
decrease, and the penalty term may become more negative. The key idea to handle both is to consider the \emph{average} over choices of $z$ on the circle $|z| =r$:
\begin{itemize}
\item 
Fix another correlation $j\neq i$. Then, the
integral of $u_z$ against $\cos(w_j s)$ can be viewed as an analytic function of
$z$. For $z$ chosen uniformly on $|z|=r$, Jensen's formula for
analytic functions gives
\[
 \mathbb E_z\!\left[
   \log\left|\int_0^{2\pi}u_z(s)\cos(w_j s)\,ds\right|
 \right]
 \geq
 \log\left|\int_0^{2\pi}u(s)\cos(w_j s)\,ds\right|.
\]
The direction of this inequality comes from the mean-value property:
if the analytic function has no zeros in the disc, the logarithm
of its absolute value has circle average equal to its value at the
center. Zeros inside the circle contribute nonnegative additional
terms in Jensen's formula, giving the lower bound above.
Note that the right-hand side is the original $j$-th term in the objective,
since $u_0=u$. Thus, every other frequency contributes no
loss to the objective on average! 

\item It remains to estimate the change in the penalty term. Recall
that this change is
\[
 CN\int_0^{2\pi}
 \bigl[\log(1-|u_z(s)|^2)-\log(1-|u(s)|^2)\bigr]\,ds.
\]
What we show is that, on average over $z$ with $|z| = r$, the loss is $O(Nr^2)$. The key observation is that
opposite choices $z$ and $-z$ produce opposite first-order changes.
These cancel when we
average over the circle. It turns out that this average can be calculated exactly (though we postpone this calculation to Section~\ref{sec:analysis_packing}): at almost every
time $s$,
\[
 \mathbb E_z\!\left[\log(1-|u_z(s)|^2)\right]
 -\log(1-|u(s)|^2)=\log(1-r^2).
\]

This change is independent of $u(s)$, so it remains controlled even
when $|u(s)|$ is close to one. Since
$-\log(1-r^2)=r^2+O(r^4)$, the average loss at each time is $O(r^2)$. Accounting for the penalty's weight $CN$ gives a loss of $O(Nr^2)$.
\end{itemize}

Putting things together, if the target $i$-th
correlation is smaller than a sufficiently small constant times $r$, the gain from our modification, on average over $z$ chosen uniformly on $|z|=r$, is
\[
 \mathbb E_z\!\left[\tilde J(u_z)-\tilde J(u)\right]
 \geq \log 2-O(Nr^2).
\]
(where recall that the $\log 2$ comes from doubling the small $i$-th correlation).
We can now see how the scale $1/\sqrt N$ appears: taking
$r=c/\sqrt N$ for a sufficiently small constant $c>0$
makes the penalty loss smaller than $\log 2$. So, there must exist some $u_z$ that has a larger objective than $u$,
contradicting maximality. Thus, every correlation must have magnitude $\Omega(1/\sqrt N)$ at the maximizer $u$.

\paragraph{Turning this function into a physical control.}
The arguments so far give a function $u$ solving the simultaneous correlation problem of \eqref{eq:overview-correlation}, which allows $u$ to be a bounded complex-valued function. Recall that this problem was a relaxation of what we arrived at via the sensing problem, which actually requires a \emph{piecewise-constant} function taking values in $\pm 1$. Recall that the latter restriction is necessary and sufficient for the function to lead to a physically implementable control.
There are two steps to turn $u$ into what we need:
\begin{itemize}
\item The first step is simply to use the fact that the squared magnitude of a complex
number is the sum of the squares of its real and imaginary parts.
Write $h_0=\textnormal{Re}(u)$ and $h_1=\textnormal{Im}(u)$.
Since cosine is real, for each candidate frequency $w_i$, at least
one of $h_0,h_1$ has a correlation with $\cos(w_i s)$ whose magnitude
is at least $1/\sqrt2$ times that of $u$. Choosing uniformly between
the two therefore gives at least $1/2$ probability of selecting a
useful one for each fixed candidate.
\item To make either real function $\pm1$-valued, the trick is to use the fact that on a short interval a cosine is
nearly constant, so its integral against either function $h_b$ depends mainly on
the average of $h_b$ there. This suggests replacing $h_b$ on a short interval by a function $f_b$ that only takes $\pm1$ values but has the same average on that interval as $h_b$ (for example, an average of $1/2$
is reproduced by spending three quarters of the interval at $+1$
and one quarter at $-1$). The approximation error can be made arbitrarily small by making the intervals smaller. 
The number of intervals needed for a sufficiently good approximation
determines the number of sign switches in the final function. In turn,
this determines how frequently control unitaries must be applied in
our protocol. In the end, our protocol needs
$O(\omega_{\max}T\sqrt N)$ control unitaries over sensing time $T$, which corresponds to a $O(\omega_{\max}\sqrt N)$ rate, where
$\omega_{\max}=\max_i\omega_i$.
\end{itemize}

The reader may finally wonder: how does one actually compute the required $u$, in order to run the protocol? In Section~\ref{sec:algorithmic}, we give a randomized algorithm that finds a suitable $u$ (not necessarily the maximizer of $J$) satisfying the desired correlations from Theorem~\ref{thm:main} in expected time $O(N^3 W \log (1 / W))$.

We conclude this overview by returning briefly to the fact that we ignored any phase $\phi$ in $H_{sig}(t) = \cos(\omega t + \phi)Z$. What we show is that an unknown phase $\phi$ can be handled in a similar way as in conventional sensing protocols, by introducing a short random wait at the start of the~protocol.

\section{Description of the sensing protocol}
\label{sec:protocol}

We are given a signal strength $B>0$ and a finite set of candidate
frequencies $\{\omega_1,\ldots,\omega_N\}\subset(0,\infty)$.
The signal Hamiltonian is promised to be either identically zero or
$$H_{\mathrm{sig}}(t)=B\cos(\omega t+\phi)Z\,,$$ where the frequency
$\omega$ belongs to the given set and the phase $\phi$ is arbitrary
and unknown.
Our goal is to distinguish these two cases with success probability
at least $2/3$, using a single sensing qubit. We allow instantaneous
single-qubit gates and count \emph{sensing time}, i.e.\ all time spent evolving under the
signal, including the random wait time at the start of each ``trial''. We describe our protocol below.
\begin{center}
\begingroup
\setlength{\fboxsep}{10pt}
\setlength{\fboxrule}{0.5pt}
\fcolorbox{blue!35!black}{blue!2}{%
\begin{minipage}{\dimexpr\linewidth-2\fboxsep-2\fboxrule\relax}
\normalsize
\refstepcounter{protocol}\label{prot:main}%
{\normalsize\textbf{Protocol~\theprotocol. Single-qubit sensing of discrete frequencies}}
\par\smallskip
{\color{blue!35!black}\hrule height 0.5pt}
\medskip
\textbf{Input.} A set of possible frequencies $\{w_1, \ldots, w_N\}$. A lower bound $B_{min} \leq B$.

\medskip
\textbf{Parameters.}
Let $\gamma>0$ be the absolute constant in
Theorem~\ref{thm:sign-bank}, and set
\[
 c:=1/\gamma,\qquad T:=c\frac{\sqrt N}{B_{min}},\qquad
 D:=\frac{2\pi}{\omega_{min}},\qquad r:=28 \,,
\]
where $\omega_{min} = \min_i\omega_i$.

\medskip
\textbf{Classical preprocessing.} Fix a pair $f_0,f_1:[0,T]\to\{-1,1\}$ from Theorem~\ref{thm:sign-bank}\footnote{In Section~\ref{sec:algorithmic}, we describe a randomized algorithm that computes a function $u$ as in Theorem~\ref{thm:main} in expected time $O(N^3 W \log (1 / W))$. Such a $u$ can then be converted into the desired pair $f_0, f_1$ following the constructive procedure in the proof of Theorem~\ref{thm:sign-bank}.}. These are
piecewise constant functions with finitely many pieces and depend
only on the known parameters. Fix both before running any trial.

\medskip
\textbf{One trial.} Perform the following steps, repeating with fresh
randomness and a newly prepared state for $r$ trials.
\begin{enumerate}[leftmargin=*,itemsep=5pt,topsep=5pt]
 \item \textbf{Sample a function.}
       Sample $b\sim\operatorname{Unif}\{0,1\}$. For the selected $f_b$,
       Theorem~\ref{thm:sign-bank} guarantees that there are $t_k$, $\sigma_k$ as follows:
       \[
        0=t_0<t_1<\cdots<t_m=T,\qquad
        f_b(t)=\sigma_k\in\{-1,1\}\quad(t_{k-1}\leq t<t_k).
       \]
       Additionally assume that empty pieces are discarded and adjacent pieces with equal
       signs are merged. Then $t_1,\ldots,t_{m-1}$ are exactly the sign-change
       times, and $\sigma_{k+1}=-\sigma_k$.
 \item \textbf{Wait, and prepare the starting state.}
       Sample a random wait time
       $\tau\sim\operatorname{Unif}[0,D]$. Let $G_0=e^{i\sigma_1\pi Y/4}$. Prepare the starting state $G_0 \ket{0}$.
 \item \textbf{Alternate signal evolution and pulses.}
       For $k=1,\ldots,m$: 
       \begin{itemize}
       \item Evolve under the signal alone for time
       $t_k-t_{k-1}$. 
       \item If $k<m$, apply
       $G_k=Y$. Thus each sign change corresponds to one $Y$ pulse.
       \end{itemize}
 \item \textbf{Undo the control and measure.}
       At time $T+\tau$, apply the inverse of the product of the
       previous gates,
       \[
        (G_{m-1}\cdots G_0)^\dagger
        =e^{-i\sigma_1\pi Y/4}Y^{m-1},
       \]
       and measure in the standard basis, recording the outcome.
\end{enumerate}
\textbf{Decision.} Output \emph{signal present} if any trial produces
outcome $1$; otherwise output \emph{signal~absent}.
\end{minipage}%
}
\endgroup
\end{center}

\paragraph{Success probability.}
When the signal is absent, the final gate undoes all the preceding
control gates, so every trial returns outcome $0$ with probability $1$. When the signal is
present, Section~\ref{sec:random-delay} shows that each trial returns
$1$ with probability at least $1/(8\pi)$, averaged over $b$ and
$\tau$, for every candidate frequency and every phase. The phase
continues to advance between trials: a trial starting at absolute
time $s$ sees phase $\phi+\omega s$. Since the bound holds for every
phase and each trial uses fresh randomness, it also holds
conditioned on all previous trials. Consequently,
\[
 \Pr[\text{all $r$ outcomes are $0$}\mid\text{signal present}]
 \leq\left(1-\frac1{8\pi}\right)^r
 \leq e^{-r/(8\pi)}<\frac13
 \qquad(r=28).
\]
Thus the protocol has no false positives and detects a present signal with probability at least~$2/3$.

\paragraph{Sensing time and number of pulses.} 
Each trial takes at most $T+D$ time. Including all waits, the total
is therefore at most
\[
 28\left(\frac{\sqrt N}{\gamma B}
          +\frac{2\pi}{\omega_{\min}}\right)
 =O\!\left(\frac{\sqrt N}{B}+\frac1{\omega_{\min}}\right) \,.
\]
In particular, this is $O(\sqrt N/B)$ when
$\omega_{\min}=\Omega(B/\sqrt N)$ (which is the typical regime). 

For the number of pulses, let $W=\max_i\omega_i$. Theorem~\ref{thm:sign-bank}
tells us that the functions $f_0$ and $f_1$ can be taken to each have at most $K = O(WT\sqrt{N}) = O(WN/B)$ sign changes. Thus, each trial uses at most that many pulses.

Sections~\ref{sec:analysis_one}, \ref{sec:analysis_packing}, \ref{sec:random-delay}, \ref{sec:analysis-four} are dedicated to proving correctness of Protocol~\ref{prot:main}.

\section{Analysis, part 1: Reducing the sensing problem to a ``simultaneous correlation'' problem}
\label{sec:analysis_one}
In our analysis, we will start from a general formulation of the problem of designing a one-qubit sensing protocol, and eventually arrive at Protocol~\ref{prot:main}. In this formulation, a sensing protocol is described by a single-qubit \emph{control} Hamiltonian $H_c(t)$, which we get to choose. Each ``trial'' begins by preparing \(\ket0\) and waiting for a time \(\tau\), after which we apply the chosen control for a duration \(T\). Throughout the analysis, \(t\in[0,T]\) denotes elapsed time after this wait, and \(\phi\) denotes the signal phase at the beginning of the trial. The signal Hamiltonian during the controlled evolution is therefore
\[
H_{\mathrm{sig}}(t)=B\cos\bigl(\omega(t+\tau)+\phi\bigr)Z.
\]
The protocol may consist of several trials, and the final decision is derived from classical post-processing of all of the measurement outcomes. For this section, we treat \(\tau\) as fixed. However, note that in Protocol~\ref{prot:main} $\tau$ is chosen uniformly at random. We will analyze the effect of the latter in Section~\(\ref{sec:random-delay}\). The primary goal in the design of the sensing protocol is to minimize total sensing time (i.e.\ Hamiltonian evolution time), and, more broadly, to do so in a way that is compatible with what can be carried out experimentally. 

In this section, we show that the problem of designing the desired sensing protocol reduces to a certain packing problem, which we will then tackle in Section~\ref{sec:analysis_packing}.

\paragraph{The control-only evolution $U_c(t)$.}

We introduce \(U_c(t)\) to denote the evolution produced by the control Hamiltonian alone. It is defined by

$$i\frac{d}{dt}U_c(t)=H_c(t)U_c(t),
\qquad U_c(0)=I \,.$$

With this, we can define a basis that ``follows'' the control. Let \(\{|0\rangle,|1\rangle\}\) be the standard basis. Define

\[
|0_c(t)\rangle:=U_c(t)|0\rangle,
\qquad
|1_c(t)\rangle:=U_c(t)|1\rangle.
\]

These basis states undergo exactly the evolution that would occur if the signal Hamiltonian were absent. Now expand the actual physical state as
\[
|\psi(t)\rangle
=
a(t)|0_c(t)\rangle+b(t)|1_c(t)\rangle.
\]
Define the coefficient vector

\[
|c(t)\rangle
:=
a(t)|0\rangle+b(t)|1\rangle
=
\begin{pmatrix}a(t)\\ b(t)\end{pmatrix}.
\]
Then the previous expansion can be written compactly as
\begin{equation}
\label{eq:1}
|\psi(t)\rangle=U_c(t)|c(t)\rangle.
\end{equation}

\paragraph{Deriving the evolution equation for $\ket{c(t)}$.}
We write $\frac{d}{dt}|\psi(t)\rangle$ in two ways: differentiating the two sides of Equation~\eqref{eq:1} and using Schrodinger's equation for $\ket{\psi(t)}$. The former gives
\[
\frac{d}{dt}|\psi(t)\rangle
=
\dot U_c(t)|c(t)\rangle
+
U_c(t)|\dot c(t)\rangle \,,
\]
and the latter gives
\[
i\frac{d}{dt}|\psi(t)\rangle
=
\big(H_c(t)+H_{\mathrm{sig}}(t)\big)|\psi(t)\rangle.
\]
Therefore, we have
\begin{equation}
\label{eq:2}
i\dot U_c(t)|c(t)\rangle
+
iU_c(t)|\dot c(t)\rangle = \big(H_c(t)+H_{\mathrm{sig}}(t)\big)
U_c(t)|c(t)\rangle \,.
\end{equation}
Now, by definition,
\[
i\dot U_c(t)\ket{v_0}=H_c(t)U_c(t)\ket{v_0}
\]
for all $\ket{v_0}$. This implies 
\[
i\dot U_c(t) =H_c(t)U_c(t) \,.
\]
Substituting this into \eqref{eq:2} gives
\[
H_c(t)U_c(t)|c(t)\rangle
+
iU_c(t)|\dot c(t)\rangle =
H_c(t)U_c(t)|c(t)\rangle
+
H_{\mathrm{sig}}(t)U_c(t)|c(t)\rangle  \,.
\]
The \(H_c\) terms cancel, and we get
\[
iU_c(t)|\dot c(t)\rangle
=
H_{\mathrm{sig}}(t)U_c(t)|c(t)\rangle.
\]
Multiplying on the left by \(U_c(t)^\dagger\) gives
\[
i|\dot c(t)\rangle
=
U_c(t)^\dagger H_{\mathrm{sig}}(t)U_c(t)|c(t)\rangle \,.
\]
Since $H_{\mathrm{sig}}(t)
=
B\cos(\omega (t + \tau) +\phi)Z$, 
we obtain
\begin{equation}
\label{eq:ct}
i|\dot c(t)\rangle
=
B\cos(\omega (t + \tau)+\phi)\,M(t)|c(t)\rangle,
\end{equation}
where
\[
M(t):=U_c(t)^\dagger ZU_c(t).
\]
Provided $[M(t),M(t')]=0$ for every pair of times $t,t'$, there is a compact expression for $\ket{c(T)}$ for any time $T$. 
The expression is
\begin{align}
\ket{c(T)}
 &=\exp\!\left(-iB\int_0^T\cos(\omega (t + \tau)+\phi)M(t)\,dt\right)
     \ket{c(0)}\nonumber\\
 &=\exp\!\left(-iB\int_0^T\cos(\omega (t + \tau)+\phi)M(t)\,dt\right)
     \ket0 \,,\label{eq:4}
\end{align}
Here $t=0$ denotes the end of the initial wait, immediately
before signal evolution or control begins. So, $\ket{c(0)}=\ket0$.

\paragraph{A special class of control Hamiltonians.} 
In our protocol, we make a special choice of control Hamiltonian: we choose $M(t)$ so that it has the simple form $M(t) = f(t)X$ where $f(t) \in \{\pm 1\}$ for all $t$. Note that, since $M(t) = U_c(t)^\dagger ZU_c(t)$ and $U_c(t)$ is unitary, $M(t)$ has the same spectrum as $Z$ (i.e.\ $\pm1$), which forces $f(t) \in \{\pm1\}$. One can check by a direct calculation that we can achieve the desired $f(t)$ by picking $U_c(t) = e^{i\theta(t)Y/2}$, where 
$$\begin{cases}
\theta(t) = \frac{\pi}{2} \quad \text{if } f(t) = 1 \\
\theta(t) = -\frac{\pi}{2} \quad \text{if } f(t) = -1
\end{cases} 
$$
The control Hamiltonian $H_c(t)$ that corresponds to such a $U_c(t)$ is proportional to $Y$, with a scalar factor consisting of a sum of delta functions. These delta functions pick out the times at which $U_c(t)$ changes from $e^{i\theta(t)Y/2}$ to $e^{-i\theta(t)Y/2}$ and viceversa. Going forward, we will simply  think of our control actions as a finite sequence of (instantaneous) unitaries, and we will focus on directly specifiying this sequence (as in Protocol~\ref{prot:main}), rather than indirectly designing $H_c(t)$.

\paragraph{The design of the sensing protocol reduces to a ``simultaneous correlation'' problem.}

Recall that in our sensing problem, we are given a known set of $N$ possible frequencies $\{\omega_1, \ldots, \omega_N\}$. Starting from $\ket{0}$, our protocol runs a Hamiltonian evolution (signal and control) for some time $T = c \cdot \frac{\sqrt{N}}{B}$ (where $c$ is defined in Protocol~\ref{prot:main}), and finally measures.

When $H_{\mathrm{sig}}(t) = 0$, the state evolves only under the control, and thus stays constant in the control-following basis. Thus, $\ket{c(T)}= \ket{0}$ for all $T$. Let
\begin{equation}
\label{eq:5}
A_{\omega,\phi,T,\tau}(f)
:=B\int_0^T f(t)\cos\bigl(\omega(t+\tau)+\phi\bigr)\,dt \,.
\end{equation}
Then, substituting $M(t)=f(t)X$ in
\eqref{eq:4} gives
\begin{equation}
 \ket{c(T)} =\cos(A_{\omega,\phi, T, \tau}(f))\ket0
       -i\sin(A_{\omega,\phi, T, \tau}(f))\ket1.
 \label{eq:exact-readout-part1}
\end{equation}
In order to detect a non-zero $H_{\mathrm{sig}}(t)$ with constant probability, we must choose $f(t)$ and the waiting time $\tau$, possibly using randomness, so that the probability of measuring $\ket{1}$ is bounded below by a constant. Equivalently, we need
\[
 \mathbb E\!\left[
   \sin^2\!\bigl(A_{\omega,\phi,T,\tau}(f)\bigr)
 \right]=\Omega(1)
\]
for every $\omega\in\{\omega_1,\ldots,\omega_N\}$ and every $\phi$,
where the expectation is over the protocol's random choices. We show that the crux is to solve the following ``simultaneous correlation'' problem, which is a relaxed version of the original one, corresponding to
$\phi=\tau=0$ and without the restriction of $f$ being  $\pm1$-valued. The problem is: find a measurable function $f:[0,T]\to\bD$, where
$\bD$ is the complex unit disc, such that
\begin{equation}
\label{eq:problem}
 \left|\int_0^T f(t)\cos(\omega_i t)\,dt\right|
 =\Omega\!\left(\frac{T}{\sqrt N}\right)
 \qquad (1\leq i\leq N).
\end{equation}
Note that, for our choice of $T=c\sqrt N/B$, the above would give $|A_{\omega_i,0,T,0}(f)|= \Omega(1)$ for all $i$, i.e.\
a constant lower bound for every candidate frequency when $\phi = \tau = 0$.

To justify the reduction to the above problem, Section~\ref{sec:sign-packing}
shows how to replace a complex-valued solution $f$ solving the problem with a \emph{pair} of
piecewise-constant functions $f_0,f_1$ taking values in
$\{-1,1\}$ with the following guarantee: for each candidate frequency $\omega_i$, at least one of the two satisfies the bound~\eqref{eq:problem} up to a constant factor. We establish this in Theorem~\ref{thm:sign-bank}. Then, our protocol, which chooses $f_0$ and $f_1$ uniformly, picks a function with the required correlation with probability
at least $1/2$. The argument in Section~\ref{sec:random-delay} shows that this
correlation guarantee also suffices when the phase $\phi$ is unknown. Specifically, recall that Protocol~\ref{prot:main} chooses $\tau$ independently and uniformly from $[0,D]$, where $D=2\pi/\min_i\omega_i$. We show in Section~\ref{sec:random-delay} that, when
$T=c\sqrt N/B$ for a sufficiently large universal constant $c$, we have
\begin{equation}
\label{eq:sine-detect}
 \mathbb{E}_{\substack{b\leftarrow\{0,1\}\\ \tau\leftarrow[0,D]}}
 \left[\sin^2\!\bigl(A_{\omega_i,\phi,T,\tau}(f_b)\bigr)\right]
 \geq\frac1{8\pi}
 \qquad (1\leq i\leq N)
\end{equation}
for every $\phi$. 

To put the problem in \eqref{eq:problem} on a fixed interval, we make the change of variables
$s=2\pi t/T$ and define
\[
 u(s):=f\!\left(\frac{T}{2\pi}s\right),
 \qquad
 \widetilde\omega_i:=\frac{\omega_iT}{2\pi}.
\]
Then,
\[
 \int_0^T f(t)\cos(\omega_i t)\,dt
 =\frac{T}{2\pi}\int_0^{2\pi}
       u(s)\cos(\widetilde\omega_i s)\,ds.
\]
Taking $T = c \sqrt{N}/B$, as in our protocol, the problem reduces to finding a measurable function $u: [0,2\pi] \rightarrow \bD$ such that
\begin{equation}
 \label{eq:goal-2}
\left|\int_0^{2\pi}u(s)\cos(\widetilde\omega_i s)\,ds\right|
 =\Omega\!\left(\frac1{\sqrt N}\right)
 \qquad (1\leq i\leq N).
\end{equation}
The rest of the proof is organized as follows. In Section~\ref{sec:analysis_packing}, we show our main theorem, which is that such a function $u$ exists (Theorem~\ref{thm:main}). In Section~\ref{sec:sign-packing}, we show how to upgrade $u$ to a pair of piecewise-constant $\pm1$-valued functions. In Section~\ref{sec:random-delay}, we show how to deal with an unknown phase $\phi$, and, in particular, that the random delay $\tau$ in our protocol yields the desired lower bound of Equation~\eqref{eq:sine-detect} on the detection probability. In Section~\ref{sec:analysis-four}, we show that the sequence of control unitaries prescribed in Protocol~\ref{prot:main} implements the desired evolution corresponding to the pair $f_0, f_1$. This completes the proof of the correctness of Protocol~\ref{prot:main}. Section~\ref{sec:discrete-lower-bound} establishes the discrete Grover-Heisenberg limit, i.e.\ a sensing time lower bound for all protocols, which matches the sensing time of our protocol asymptotically.

\section{Analysis, part 2: Constructing a bounded function with large correlations}
\label{sec:analysis_packing}
In this section, we study the key technical problem that underlies our sensing protocol.

Let $\bD_r = \{z \in \C: |z| < r\}$ denote the open complex disc of radius $r$, and let $\bD = \bD_1$ denote the unit disc.
Let $\bT$ denote the set of $L^2$ integrable functions that map $[0, 2\pi] \to \bD$. We show the following.
\begin{theorem}
\label{thm:main}
    There exists a universal constant $c > 0$ so that for every integer $N \geq 1$ and any $\omega_1, \ldots, \omega_N > 0$, there is a function $u \in \bT$ such that, for all $i = 1, \ldots, N$, we have
    $$
    \left| \int_0^{2 \pi} u(x) \cos (\omega_i x) dx \right| \geq \frac{c}{\sqrt{N}} \; .
    $$
\end{theorem}
This is the key theorem that powers our sensing protocol. In Section~\ref{sec:sign-packing}, we will upgrade this theorem so that it guarantees a \emph{pair} of real-valued functions such that, for each $i$, at least one of the two functions has the desired overlap with $\cos(\omega_i t)$. Then, we will upgrade the theorem further so that it guarantees a pair of $\pm1$-valued functions (those eventually used in Protocol~\ref{prot:main}). Finally, in Section~\ref{sec:random-delay}, we will show that the random delay $\tau$ in Protocol~\ref{prot:main} handles the unknown phase~$\phi$.

We first need the following facts from complex analysis.
\begin{theorem}[Jensen's formula]
\label{thm:jensen}
    Let $r>0$. Let $F$ be holomorphic on an open neighborhood of the closed disc $\overline{\bD_r}$ with $F(0) \neq 0$, and let $\{\zeta_j\}$ denote all the roots of $F$ satisfying $|\zeta_j| < r$, counting multiplicity. 
    Then,
    $$
    \frac{1}{2 \pi} \int_0^{2 \pi} \log |F(r e^{i \theta})| d \theta = \log |F(0)| + \sum_j \log \frac{r}{|\zeta_j|} \; .
    $$
    (Here, we allow $F$ to have zeros on the boundary since the logarithmic singularities are integrable).
\end{theorem}

\begin{corollary}
\label{cor:log-integral-nonneg}
    Let $r>0$ and let $g$ be holomorphic on an open neighborhood
of the closed disc $\overline{\bD_r}$. Then,
$$
 \frac{1}{2\pi}\int_0^{2\pi}\log|g(re^{i\theta})|\,d\theta
 \geq \log|g(0)| \,,
$$
where $\log0=-\infty$.
\end{corollary}
\begin{proof}
If $g(0)=0$, the inequality is immediate. Otherwise, the inequality follows directly from Theorem~\ref{thm:jensen} and the fact that, for each zero $\zeta_j$ inside the disc, 
$\log(r/|\zeta_j|)\geq0$.
\end{proof}

\begin{theorem}[Rouch\'{e}'s theorem]
    Suppose $f, g$ are two holomorphic functions on an open neighborhood of the closed disc $\overline{\bD_r}$, and suppose that $|g(z)| < |f(z)|$ for all $|z| = r$.
    Then $f$ and $f + g$ have the same number of zeros (counting multiplicity) on the open disc $\bD_r$.
\end{theorem}

We will also need the following family of very useful conformal maps.
For any $a$ satisfying $|a| < 1$, define $L_{\pm, a} (z)$ by
\[
L_{\pm, a} (z) = \frac{a \pm z}{1 \pm \bar{a} z} \; .
\]
The following standard properties of these so-called \emph{M\"{o}bius transformations} or linear fractional transformations will be very useful for us, and can be straightforwardly verified:
\begin{fact}
\label{fact:55}
    For all $a$ satisfying $|a| < 1$, we have:
    \begin{itemize}
        \item $L_{\pm, a}$ is a holomorphic function mapping $\bD$ to $\bD$ satisfying $L_{\pm, a} (0) = a$.
        \item $|L_{\pm, a} (z)| \leq 1$ for all $|z| \leq 1$, with equality if and only if $|z| = 1$.
        \item We have:
        \[
        1 - \left| L_{\pm, a} (z) \right|^2 = \frac{(1 - |a|^2) (1 - |z|^2)}{|1 \pm \bar{a} z|^2} \; .
        \]
        \item $L_{\pm, a}$ has the following series expansion around $z = 0$, valid for all $|z| < 1$:
        \[
        L_{\pm, a} (z) = a + (1 - |a|^2) \sum_{\ell = 1}^\infty (\pm 1)^\ell (-\bar{a})^{\ell - 1} z^\ell \; .
        \]
    \end{itemize}
\end{fact}
\begin{proof}
    The first three bullets follow from the fact that $1\pm\bar a z$ is non-zero on the closed unit disc (and a direct calculation for the third fact).
    To see the last one, we first subtract $a$, since $L_{\pm, a} (0) = a$, to obtain that
    \[
    L_{\pm, a}(z) - a = \frac{\pm ( 1 - |a|^2) z}{1 \pm \bar{a} z} \; ,
    \]
    and then we plug in the standard series expansion $\frac{1}{1-z} = \sum_{m = 0}^\infty z^m$, which is valid for all $|z| < 1$.
\end{proof}

We will need the following straightforward bound:
\begin{lemma}
\label{lem:packing-cos-bound}
    For all $\omega > 0$, we have $\int_{0}^{2 \pi} |\cos (\omega x)| dx \geq \pi - 1$.
\end{lemma}
\begin{proof}
    Since $|\cos(\omega x)| \geq \cos^2(\omega x)$, we have
    \[
    \int_0^{2\pi} |\cos(\omega x)|\,dx
    \geq \int_0^{2\pi} \cos^2(\omega x)\,dx
    = \pi + \frac{\sin(4\pi \omega)}{4 \omega}.
    \]
    If $0 < \omega \leq 1/4$, the right-hand side is at least $\pi$. If $\omega \geq 1/4$, it is at least $\pi - 1/(4\omega) \geq \pi - 1$.
\end{proof}

With these tools, we can now prove Theorem~\ref{thm:main}. For any two $L^2$ functions $f, g: [0, 2\pi] \to \C$, let $\iprod{f, g} = \frac{1}{2 \pi} \int_{0}^{2 \pi} \overline{f(x)} g(x) dx$, and let $\norm{f}^2 = \iprod{f, f}$. Let $g_i (x) = \cos (\omega_i x)$ for $i = 1, \ldots, N$.
The key idea will be to consider the following potential functional $J: \bT \to \R \cup \{-\infty\}$, defined by
\begin{equation}
\label{eq:J}
J(f) = \sum_{i = 1}^N \log \left| \iprod{g_i, f} \right| + C N \cdot \frac{1}{2 \pi} \int_{0}^{2 \pi} \log (1 - |f(x)|^2) dx \; .
\end{equation}
Here $C>0$ is some constant to be chosen later to be sufficiently large, and we use the convention \(\log0=-\infty\).

The first step is to establish that $J$ indeed has a maximizer.
\begin{lemma}[Existence of a maximizer]
\label{lem:packing-maximizer-exists}
For every $C>0$, the functional $J$ attains a finite maximum
on $\bT$.
\end{lemma}
To keep the focus on the crux of the argument, we postpone the proof of this to Appendix~\ref{sec:app}. 

We first establish the following bound on the behavior of $J$ under scaling.
\begin{lemma}
\label{lem:packing-energy-bound}
Suppose $u \in \bT$ satisfies $\norm{u}^2 > 1 / (2C)$.
Then, 
\[
J \left( \frac{1}{\sqrt{2C}} \frac{u}{\norm{u}} \right) \geq J(u) \; .
\]
The inequality is strict whenever $J(u)$ is finite. In particular, every maximizer $u$ of $J$ satisfies $\norm{u}^2 \leq 1 / (2C)$.
\end{lemma}
\begin{proof}
If $J(u)=-\infty$, the inequality is immediate.
Otherwise, for $0<t\leq1$, define $\gamma(t)=J(tu)$. 
Using monotone convergence to justify differentiating under the integral, we have that
\[
t \gamma' (t) = N - \frac{2CN}{2\pi}\int_0^{2\pi}
       \frac{t^2 |u(x)|^2}{1- t^2 |u(x)|^2}\,dx.
\]
Since $t^2 |u(x)|^2/(1-t^2 |u(x)|^2)\geq t^2 |u(x)|^2$, the derivative $\gamma'(t)$ is strictly negative for all $t \in \left( \frac{1}{\sqrt{2C} \norm{u}}, 1 \right)$. Also, $\gamma(t)\to J(u)$ as $t\uparrow1$. Hence, we have
$J\left(\frac{1}{\sqrt{2C}}\frac{u}{\norm{u}}\right)>J(u)
$. Every maximizer has finite value by
Lemma~\ref{lem:packing-maximizer-exists}, so if it were
$\norm{u}^2>1/(2C)$, the strict inequality would contradict maximality. 
\end{proof}
There is a fairly intuitive explanation of what is happening here.
The bound quantifies how the two terms in $J$ balance under
a shrinking map $u\mapsto(1-\varepsilon)u$. To first order in
$\varepsilon$, the sum of the logarithms of the correlations decreases by
$N\varepsilon$, while the penalty becomes less negative by at least
$2CN \norm{u}^2\varepsilon$. At a maximizer, this improvement must be
balanced by at least as much loss from the correlations, as otherwise
shrinking $u$ would increase $J$. So, it must be $2C N \norm{u}^2\leq N$,
so the constant $C$ controls how small the average squared
magnitude must be.

We now proceed to the main proof of Theorem~\ref{thm:main}.
The key technique (which will also be useful for us algorithmically) is to consider what happens to $J(f)$ when we take a suitable random holomorphic perturbation of $f$ at every point.
Formally, for any $i = 1, \ldots, N$, let $s_i (x) = \mathrm{sign} (g_i (x))$ for all $x \in [0, 2 \pi]$, and for any $z \in \bD$, define the transformation $\Phi^{(i)}_z: \bT \to \bT$ given by 
    \[
    \left( \Phi_z^{(i)} f \right) (x) = L_{s_i (x), f(x)} (z) \; .
    \]
    Notice in particular that $\Phi_0^{(i)}  f = f$ for all $i$. Since each map $L_{\pm,a}$ sends $\bD$ into itself, we have
$(\Phi_z^{(i)} f)(x)\in\bD$ for every $x$ whenever $|z|<1$.
One can see straightforwardly that the transformed function is also measurable, so $\Phi_z^{(i)} f\in\bT$.

Let $\mathbb E_{|z| = r}$ denote integrating over $z=re^{i\theta}$,
with $\theta$ from the uniform measure on $[0,2\pi]$. 

\begin{lemma}
\label{lem:potential-progress}
    Let $r = \eta / \sqrt{N}$ for $\eta \in (0, 1/16)$ and let $u \in \bT$.
    Then, for all $i, j \in [N]$, we have that 
    \[
        \mathbb E_{|z| = r}\!\left[ \log\frac{\left|\iprod{g_j,\Phi_z^{(i)} u}\right|}{|\iprod{g_j,u}|} \right]\geq0.
    \]
    Moreover, suppose $u \in \bT$ satisfies $\| u \|^2 \leq 1/(2C)$, and $0 < \left| \langle g_i, u \rangle \right| \leq c / \sqrt{N}$, for $c \leq \eta / 32$.
    Then we have that
    \[
        \mathbb E_{|z| = r}\!\left[ \log\frac{\left|\iprod{g_i,\Phi_z^{(i)} u} \right|}{|\iprod{g_i,u}|} \right]\geq \log 2.
    \]
\end{lemma}
\begin{proof}
For each $j$, the function
$z\mapsto\iprod{g_j,\Phi_z^{(i)} u}$ is clearly holomorphic on $\bD$. 
At $z=0$, its value is $\iprod{g_j, u}$, since $\Phi_0^{(i)} u=u$.
Corollary~\ref{cor:log-integral-nonneg} thus implies the first inequality.

To show the second inequality, we claim that the function $h(z) = \iprod{g_i, \Phi_z^{(i)} u}$ has a root at a point $\zeta$ satisfying $|\zeta| \leq r / 2$.
Then, by Jensen's formula, the LHS of the inequality is at least $\log(r/|\zeta|) \geq \log 2$.
To do so, the strategy is to use Rouch\'{e}'s theorem.
By the series expansion of $L$ from Fact~\ref{fact:55},
\begin{align*}
 h(z)
 &=\iprod{g_i,L_{s,u}(z)}\\
 &=\iprod{g_i,u+(1-|u|^2)sz
   +(1-|u|^2)\sum_{\ell\geq2}
       s^\ell(-\bar u)^{\ell-1}z^\ell}\\
 &=\underbrace{
   \iprod{g_i,u}
   +\frac{z}{2\pi}\int_0^{2\pi}
       (1-|u(x)|^2)|g_i(x)|\,dx
   }_{\coloneqq f(z)} \,\,+\,\,\underbrace{
   \iprod{g_i,(1-|u|^2)\sum_{\ell\geq2}
       s^\ell(-\bar u)^{\ell-1}z^\ell}
   }_{\coloneqq g(z)}.
\end{align*}
Here the last equality uses $g_i(x)s(x)=|g_i(x)|$. We will apply Rouch\'{e}'s theorem on the circle $|z|=r/2$.
To do so, we show that $f$ has a zero inside this circle and
that $|g(z)|<|f(z)|$ everywhere on the circle. Here is the point: $f$ is linear, with constant term
$\iprod{g_i,u}$ and a slope bounded below by a constant, by Lemma~\ref{lem:packing-cos-bound}, our assumption on $\| u \|^2$, and our choice of $C$.
If $|\iprod{g_i,u}|$ is sufficiently small compared to $r$,
the zero of $f$ must lie close to the origin. Moreover $|f(z)|$ is $\Theta(r)$ on $|z|=r/2$.
On the other hand, the series for $g$ starts with $z^2$, and we will
bound $|g(z)|$ by $O(r^2)$ on the same circle. For sufficiently small $r$, this is smaller than $|f(z)|$, which is the required bound for Rouch\'e's
theorem. We now make this precise.

First, Lemmas~\ref{lem:packing-cos-bound}
and~\ref{lem:packing-energy-bound} give the following lower
bound on the coefficient of $z$ in $f$:
\begin{align*}
 \frac1{2\pi}\int_0^{2\pi}(1-|u(x)|^2)|g_i(x)|\,dx
 &\geq \frac{\pi-1}{2\pi}-\norm{u}^2\\
 &\geq \frac{\pi-1}{2\pi}-\frac1{2C}
 >\frac18,
\end{align*}
where we used that $C\geq4$.

Recall that $r=\eta/\sqrt N$. Choosing $c\leq\eta/32$,
our assumption on the original correlation gives
\[
 |\iprod{g_i,u}|\leq\frac{c}{\sqrt N}\leq\frac r{32}.
\]
Since $f$ is linear and its slope is at least $1/8$, its unique
zero has magnitude at most
$8|\iprod{g_i,u}|\leq r/4$. Thus this zero lies inside
$|z|<r/2$. On the boundary $|z|=r/2$, a triangle
inequality also gives
\[
 |f(z)|
 \geq\frac18|z|-|\iprod{g_i,u}|
 \geq\frac r{16}-\frac r{32}
 =\frac r{32}.
\]

It remains to show that the remainder $g$ is smaller than this
on the same circle. Using the triangle inequality, and $|g_i(x)|\leq1$, $|u(x)|<1$, we have
\begin{align*}
 |g(z)|
 &\leq\frac1{2\pi}\int_0^{2\pi}
 |g_i(x)|(1-|u(x)|^2)
 \sum_{\ell\geq2}|u(x)|^{\ell-1}|z|^\ell\,dx\\
 &\leq\sum_{\ell\geq2}|z|^\ell
 =\frac{|z|^2}{1-|z|}.
\end{align*}
The key point is that $|g(z)|$ is quadratic in $|z|$, whereas our lower
bound on $|f(z)|$ is linear in $r$. So, we can make  $|g(z)|$ smaller by choosing $\eta$, and hence $r$, sufficiently small.
Choosing $\eta\leq1/16$ makes $r\leq1/16$, and, on $|z|=r/2$, we have
\[
 |g(z)|
 \leq\frac{(r/2)^2}{1-r/2}
 <\frac{r^2}{2}
 \leq\frac r{32}
 \leq|f(z)|.
\]

Both $f$ and $g$ are holomorphic on $\bD$, so Rouch\'e's theorem
applies on $|z|=r/2$. It follows that $h=f+g$ has the same number
of zeros as $f$ inside this circle, namely one. Hence, $h$ has
a zero $\zeta$ with $|\zeta|<r/2$, proving the claim.
\end{proof}

\noindent
This yields the following corollary:
\begin{corollary}
\label{cor:potential-progress}
    Let $u \in \bT$ be so that $\norm{u}^2 \leq 1 / (2C)$ and $|\iprod{g_i, u}| < c / \sqrt{N}$ for some $i$.
    Let $r = \eta / \sqrt{N}$, where $\eta$ satisfies $4C\eta^2<\log2$, and $c$ satisfies $c \leq \eta / 32$.
    Then, we have that
    \begin{equation}
    \label{eq:potential-progress}
    \mathbb{E}_{|z| = r} \left[ J(\Phi_z^{(i)} (u)) - J(u) \right] \geq \frac{1}{2} \log 2 \; .
    \end{equation}
\end{corollary}
\begin{proof}
        For the sake of conciseness, since we will only use the index $i$ in this proof, we will drop it and refer to $\Phi_z^{(i)}$ as $\Phi_z$, and similarly $s_i$ as $s$.
        By Lemma~\ref{lem:potential-progress}, we have that
    \begin{align*}
        \mathbb{E}_{|z| = r} \left[ J(\Phi_z (u)) - J(u) \right] &= \mathbb{E}_{|z| = r} \left[ \sum_{j = 1}^N \log \frac{|\iprod{g_j, \Phi_z  u}|}{|\iprod{g_j, u}|} + CN \cdot \frac{1}{2 \pi} \int_{0}^{2\pi} \log \frac{1 - |\Phi_z (u) (x) |^2}{1 - |u(x)|^2} dx \right] \\
        &\geq \log 2 + CN \cdot \mathbb{E}_{|z| = r} \left[ \frac{1}{2 \pi} \int_{0}^{2\pi} \log \frac{1 - |\Phi_z (u) (x) |^2}{1 - |u(x)|^2} dx \right] \; .
    \end{align*}
    By Fact~\ref{fact:55}, we can simplify the expression in the expectation as
    \[
    \mathbb{E}_{|z| = r} \left[ \frac{1}{2 \pi} \int_{0}^{2\pi} \log \frac{1 - |\Phi_z (u) (x) |^2}{1 - |u(x)|^2} dx \right] =  \frac{1}{2 \pi} \int_{0}^{2 \pi} \mathbb{E}_{|z| = r} \left[  \log \frac{1 - |z|^2}{|1 + s(x) \overline{u(x)} z|^2}  \right] dx \; .
    \]
    The function $u(z) = 1 + s(x) \overline{u(x)} z$ has no zeros on the unit disc for all $x$, and hence by Jensen's formula, we have
    \begin{align*}
        \frac{1}{2 \pi} \int_{0}^{2 \pi} \mathbb{E}_{|z| = r} \left[  \log \frac{1 - |z|^2}{|1 + s(x) \overline{u(x)} z|^2}  \right] dx &= \log (1 - r^2) - \frac{1}{2 \pi} \int_{0}^{2 \pi} \mathbb{E}_{|z| = r} \left[  \log |1 + s(x) \overline{u(x)} z|^2  \right] dx \\
        &= \log (1 - r^2) \; ,
    \end{align*}
    so overall, we have that 
    \begin{align*}
    \mathbb{E}_{|z| = r} \left[ J(\Phi_z (u)) - J(u) \right] &\geq \log 2 + C N \log (1 - r^2) \\
    &\geq \log 2 - C N \frac{r^2}{1 - r^2}  \\
    &= \log 2 - CN \frac{\eta^2 / N}{1 - \eta^2 / N} \geq \frac{1}{2} \log 2 \; ,
    \end{align*}
    by our choice of $\eta$, as claimed.
\end{proof}

\begin{proof}[Proof of Theorem~\ref{thm:main}]
Let $u \in \bT$ be a maximizer of $J(u)$. By Lemma~\ref{lem:packing-maximizer-exists} this is guaranteed to exist, and, by Lemma~\ref{lem:packing-energy-bound}, it satisfies $\norm{u}^2 \leq (1 / 2C)$.
With our choice of parameters, Corollary~\ref{cor:potential-progress} implies that if $|\iprod{g_i, u}| < c / \sqrt{N}$, then there exists some $z \in \bD$ so that $J (\Phi_z^{(i)} (u)) > J(u)$, contradicting the maximality of $u$.
\end{proof}

\paragraph{An algorithmic version of Theorem~\ref{thm:main}}
\label{sec:algorithmic}

Theorem~\ref{thm:main} demonstrates that for any choice of $\omega_1, \ldots, \omega_N$, there exists a good witness function $u$.
However, \emph{a priori}, it does not yield an algorithm to find such a function.
But the proof strategy actually lends itself to a particularly natural algorithm for finding a good witness function.
Namely, Corollary~\ref{cor:potential-progress} demonstrates that a random M\"{o}bius transformation of a candidate function that does not satisfy all of the inner product conditions will increase the potential function in expectation by a constant amount.
However, the potential function is always non-positive, so this process cannot continue for too long in expectation.
Formally, we show:
\begin{theorem}
\label{thm:algorithmic}
    Let $C, r, \eta,$ and $c$ be as in the previous subsection.
    There is a universal constant $c > 0$ and an algorithm, which, given $\omega_1, \ldots, \omega_N$, outputs a $u \in \bT$ so that $|\iprod{g_i, u}| \geq c / \sqrt{N}$ for all $i \in [N]$.
    The algorithm outputs a piecewise constant function with $O(N W)$ pieces, and runs in expected time $O(N^3 W \log ( W))$, where 
    \[
    W = \max \left( \frac{C^2}{32 \pi^2}, \max_{i \in [N]} \omega_i \right) \; .
    \]
\end{theorem}

\noindent
We note that by standard tricks, for all $\delta > 0$, this also yields an algorithm which runs in time $O(N^3 W \log W \cdot \tfrac{1}{\delta})$, and which succeeds with probability at least $1 - \delta$.
This is a purely classical preprocessing step, and so for clarity of exposition, we defer the description of the algorithm and the proof of its correctness to Appendix~\ref{app:algorithmic}.

\subsection{Upgrading Theorem~\ref{thm:main} to a pair of $\pm1$-valued functions}
\label{sec:sign-packing}
The output of Theorem~\ref{thm:main} is a function that is complex-valued. In this
section, we upgrade it to a pair of $\pm1$-valued functions. The two main ideas are the following:
\begin{itemize}
  \item To obtain a real-valued function, consider the real and imaginary parts of the function $u$ supplied by Theorem~\ref{thm:main}. One of them must yield a large integral (simply because the squared contribution from $u$ is the sum of the squares of the real and imaginary contributions).
  Here, one subtlety is that, from Theorem~\ref{thm:main} we do not get a real-valued function that works for all $\omega_i$, but, rather, for all $i$, one of the two functions has the desired guarantee. So our sensing protocol eventually picks one of the two uniformly at random, incurring only a constant loss in success probability.
  \item To obtain a $\pm1$-valued function from a real-valued function, we split $[0,2\pi]$ into tiny intervals on which the cosine functions are almost constant. We then reproduce the function's average on that interval using the appropriate proportion of $+1$ and $-1$ contributions.
\end{itemize}
We start with the following lemma.
\begin{lemma}
\label{lem:binary-rounding}
Let $h:[0,T]\to[-1,1]$ be measurable, and partition $[0,T]$ into
$K$ equal intervals of length $\delta=T/K$. For each interval
$I=[a,a+\delta]$, let
\[
 p_I:=\frac{\delta+\int_I h(t)\,dt}{2},\qquad
 \mathcal R_K(h)(t):=
 \begin{cases}
 +1,&a\leq t<a+p_I,\\
 -1,&a+p_I\leq t<a+\delta.
 \end{cases}
\]
Use half-open intervals and the left limiting sign at $T$.
This gives a function taking values in $\{-1,1\}$ everywhere,
with at most $2K$ constant pieces. For any real $L$-Lipschitz
function $g$ on $[0,T]$,
\begin{equation}
 \left|\int_0^T\bigl(\mathcal R_K(h)-h\bigr)g(t)\,dt\right|
 \leq \frac{LT^2}{2K}.
 \label{eq:rounding-bound}
\end{equation}
\end{lemma}
\begin{proof}
We have $0\leq p_I\leq\delta$ and
$\int_I(\mathcal R_K(h)-h)=0$ on each interval. Let $c_I$ be the midpoint of $I$. Since
$\int_I(\mathcal R_K(h)-h)\,dt=0$, multiplying this integral
by the constant $g(c_I)$ still gives zero. Therefore,
\[
 \int_I(\mathcal R_K(h)-h)g(t)\,dt
 =
 \int_I(\mathcal R_K(h)-h)\bigl(g(t)-g(c_I)\bigr)\,dt.
\]
Now $|\mathcal R_K(h)-h|\leq 2$, and the Lipschitz assumption
gives $|g(t)-g(c_I)|\leq L|t-c_I|$. Thus
\[
 \left|\int_I(\mathcal R_K(h)-h)g(t)\,dt\right|
 \leq 2L\int_I |t-c_I|\,dt
 =\frac{L\delta^2}{2},
\]
where we used
$\int_I |t-c_I|\,dt=2\int_0^{\delta/2}u\,du=\delta^2/4$. Finally, a triangle inequality gives the desired bound \eqref{eq:rounding-bound} for the integral over
$[0,T]$.
\end{proof}

The lemma's statement uses general Lipschitz test functions because
the rounding operation depends only on $h$. In particular it gives
the same error for $g(t)=\cos(\omega t+\phi)$ for every $\phi$,
since its Lipschitz constant is at most $\omega$. This elementary
uniformity will also be used when a delay changes the phase $\phi$.

\begin{theorem}
\label{thm:sign-bank}
For every $T>0$ and every finite set
$\omega_1,\ldots,\omega_N>0$, there exist two piecewise-constant
functions $f_0,f_1:[0,T]\to\{-1,1\}$ such that
\begin{equation}
 \max_{b\in\{0,1\}}
 \left|\int_0^T f_b(t)\cos(\omega_i t)\,dt\right|
 \geq\frac{\gamma T}{\sqrt N}
 \quad(1\leq i\leq N ) \,,
 \label{eq:sign-bank}
\end{equation}
where $\gamma>0$ is a universal constant. Moreover, $f_0$ and $f_1$ can be taken to each have at most $K = \Theta(WT\sqrt{N}+1)$ sign changes, where $W=\max_i\omega_i$.
\end{theorem}
\begin{proof}
Apply Theorem~\ref{thm:main} with frequencies
$w_i= \frac{\omega_i T}{2\pi}$ to obtain a corresponding function $u$. Let
$u_T(t)=u(2\pi t/T)$. Then, a straightforward change of variables calculation gives
\[
 \left|\int_0^T u_T(t)\cos(\omega_i t)\,dt\right|
 \geq\frac{\kappa T}{\sqrt N}.
\]
where $\kappa = \frac{c}{2\pi}$, and $c$ is the constant in the statement of Theorem~\ref{thm:main}. Write $h_0= \textnormal{Re}(u_T)$ and $h_1  = \textnormal{Im}(u_T)$. Now, since the squares of real and imaginary parts sum to the square of the absolute value (and cosine is real), we have that, for every $i$, some $b$ satisfies
\[
 \left|\int_0^T h_b(t)\cos(\omega_i t)\,dt\right|
 \geq\frac{\kappa T}{\sqrt{2N}}.
\]
Let $W=\max_i\omega_i$. Let
\[
 K=\left\lceil WT\sqrt{2N}/\kappa\right\rceil,
\]
and let $f_b=\mathcal R_K(h_b)$ for $b\in\{0,1\}$.
Each $f_b$ has at most $2K-1=O(1+WT\sqrt N)$ sign changes. Indeed, there is at most one sign change inside each of the $K$
intervals and at most one at each of the $K-1$ boundaries between them. Now, since $t\mapsto\cos(\omega_i t)$ is $W$-Lipschitz,
applying Lemma~\ref{lem:binary-rounding} with $K$ intervals gives, for every $i$ and $j$,
\[
 \left|
 \int_0^T f_j(t)\cos(\omega_i t)\,dt
 -
 \int_0^T h_j(t)\cos(\omega_i t)\,dt
 \right|
 \leq \frac{WT^2}{2K}
 \leq \frac{\kappa T}{2\sqrt{2N}}.
\]
For each $i$, choose an index $b=b(i)$ such that
$\left|\int_0^T h_b(t)\cos(\omega_i t)\,dt\right|
\geq\kappa T/\sqrt{2N}$.
Then, by a triangle inequality, we have
\[
 \left|\int_0^T f_b(t)\cos(\omega_i t)\,dt\right|
 \geq \frac{\kappa T}{\sqrt{2N}}
       -\frac{\kappa T}{2\sqrt{2N}}
 =\frac{\gamma T}{\sqrt N},
\]
where $\gamma=\frac{\kappa}{2\sqrt2}$.
\end{proof}

\section{Analysis, part 3: Handling the unknown phase $\phi$}
\label{sec:random-delay}
In this section, we show that our protocol detects a signal with constant probability even in the presence of an unknown phase $\phi$.
Let
$f_0,f_1:[0,T]\to\{-1,1\}$ be the two piecewise functions supplied by Theorem~\ref{thm:sign-bank}. Then, 
for each candidate frequency $\omega_i$, at least one $b\in\{0,1\}$ satisfies
$$
 \left|\int_0^T f_b(t)\cos(\omega_i t)\,dt\right| \geq\frac{\gamma T}{\sqrt N}\,.
$$
Let $F_{i,b} = \int_0^T f_b(t)e^{i\omega_i t}\,dt$. Then, 
$$ | F_{i,b} | \geq  |\Re(F_{i,b})| = \left| \int_0^T f_b(t)\cos(\omega_i t)\,dt \right| \geq\frac{\gamma T}{\sqrt N}\,.$$
For a signal with unknown phase $\phi$, however, the relevant integral is
\[
 \int_0^T f_b(t)\cos(\omega_i t+\phi)\,dt
 =\Re\!\left(e^{i\phi}F_{i,b}\right)
 =|F_{i,b}|\cos(\phi+\arg F_{i,b}),
\]
which may vanish despite the lower bound on $|F_{i,b}|$.

To avoid this cancellation, Protocol~\ref{prot:main} includes a random waiting time $\tau\sim\operatorname{Unif}[0,D]$, where $D = \frac{2\pi}{\omega_{\min}}$ and $\omega_{\min}=\min_i\omega_i$. We now analyze its effect. From the analysis in part 1, Equation \eqref{eq:exact-readout-part1}, the probability of detecting the signal is
$P_{i,b}(T, \tau) :=\sin^2\!\bigl(A_{\omega_i,\phi, T, \tau}(f)\bigr)$, where recall that
\begin{equation}
 \label{eq:delayed-signal-integral}
 A_{\omega_i,\phi, T, \tau}(f)
 = B \int_0^T f_b(t)\cos\bigl(\omega_i(t+\tau)+\phi\bigr)\,dt
 =B |F_{i,b}|\cos\bigl(\phi+\omega_i\tau+\arg F_{i,b}\bigr).
\end{equation}

Fix a candidate frequency $\omega_i$ and an index $b\in\{0,1\}$ for
which $|F_{i,b}|\geq\gamma T/\sqrt N$. Taking
$T\geq\sqrt N/(\gamma B)$ ensures that $B|F_{i,b}|\geq1$.
First consider averaging the detection probability over any waiting-time
interval of length $2\pi/\omega_i$ (instead of averaging over $[0,D]$). Over this interval, the phase
$\phi+\omega_i\tau+\arg F_{i,b}$ goes through one full period uniformly, so this average is
\begin{align*}
 &\frac1{2\pi}\int_0^{2\pi}
       \sin^2\!\bigl(B|F_{i,b}|\cos\theta\bigr)\,d\theta\\
 &=\frac1\pi\int_{-1}^{1}
       \frac{\sin^2\!\bigl(B|F_{i,b}|x\bigr)}{\sqrt{1-x^2}}\,dx\\
 &\geq\frac1\pi\int_{-1}^{1}
       \sin^2\!\bigl(B|F_{i,b}|x\bigr)\,dx\\
 &=\frac1\pi\left(1-
       \frac{\sin(2B|F_{i,b}|)}{2B|F_{i,b}|}\right)
 \geq\frac1{2\pi}.
\end{align*}
The first equality uses the substitution $x=\cos\theta$, accounting for
the two halves of the period. The last inequality uses $B|F_{i,b}|\geq1$ and
$\sin(2B|F_{i,b}|)\leq1$.

Now consider the actual waiting interval $[0,D]$. Since
$D=2\pi/\omega_{\min}\geq2\pi/\omega_i$, it contains at least one
full period. Partition it, starting at zero, into intervals of length
$2\pi/\omega_i$ and a possible shorter remainder. The full intervals
occupy at least half of $[0,D]$. Each full interval has average detection probability at least
$1/(2\pi)$, independently of the unknown phase $\phi$.
discarding the nonnegative contribution from the remainder therefore gives
\[
 \mathbb E_\tau\bigl[P_{i,b}(T,\tau)\bigr]
 \geq\frac12\cdot\frac1{2\pi}=\frac1{4\pi}.
\]
Finally, the protocol chooses $b$ uniformly from $\{0,1\}$,
independently of $\tau$. For each candidate frequency, at least one of
the two functions $f_0,f_1$ satisfies the required bound.
So, the probability of selecting such a function is at least
$1/2$. Thus, for every candidate frequency and every phase,
\begin{equation}
 \mathbb E_{b,\tau}\bigl[P_{i,b}(T,\tau)\bigr]
 \geq\frac12\cdot\frac1{4\pi}=\frac1{8\pi}.
 \label{eq:averaged-detection-probability}
\end{equation}

\section{Analysis, part 4: Putting things together}
\label{sec:analysis-four}
In this section, we show that the desired control Hamiltonian can be implemented with an appropriately chosen \emph{finite} sequence of unitaries (i.e.\ ``instantaneous pulses''), which alternate with evolution according to the signal Hamiltonian, as described in Protocol~\ref{prot:main}.

Fix
one of the two functions $f_0,f_1$ from
Theorem~\ref{thm:sign-bank}, call it $f$, and write
\[
 0=t_0<t_1<\cdots<t_m=T,
 \qquad f(t)=\sigma_k\in\{-1,1\}
 \quad(t_{k-1}\leq t<t_k).
\]
Assume empty pieces are discarded and adjacent pieces with the same sign are merged. All the times $t_k$ and signs $\sigma_k$ are known to the protocol. As in Section~\ref{sec:analysis_one}, $t$ denotes elapsed time
after the initial wait, so the signal Hamiltonian during this
interval is $B\cos(\omega(t+\tau)+\phi)Z$.
\paragraph{The instantaneous pulses in our protocol have the desired effect.}
Let $\ket{\psi(t)}$ denote the physical state generated by the protocol.
Its coefficient vector in the control-following basis is then defined as
\[
\ket{c(t)}:=U_c(t)^\dagger\ket{\psi(t)},
\]
where $U_c(t)$ is the control-only evolution. Thus
$\ket{\psi(t)}=U_c(t)\ket{c(t)}$ holds by definition. By construction, our protocol alternates control unitaries and continuous evolution under the signal Hamiltonian. So, $U_c(t)$ is constant on an interval $(t_{k-1}, t_k)$. Let $C_k$ be such that $U_c(t) = C_k$ on the latter interval. Then, 
$$ C_k = G_{k-1} \ldots G_0 \,,$$
where the $G_j$'s are as in Protocol~\ref{prot:main}.

When no signal is present, i.e.\ $H_{sig}(t) = 0$, then $\ket{c(t)} = \ket{0}$ for all $t$.

When signal is present, we will prove that $\ket{c(t)}$ starts at $\ket0$, remains
continuous across pulses, and satisfies the equation
\begin{equation}
\label{eq:coeff-eq}
 i\frac{d}{dt}\ket{c(t)}
 =B\cos(\omega(t+\tau)+\phi)\,\sigma_kX\ket{c(t)}
\end{equation}
between pulses, for $t_{k-1}<t<t_k$ where $f(t)=\sigma_k$ for $t \in [t_{k-1}, t_k)$. Assuming this is the case, then uniqueness of the (continuous) solution to the Schrodinger equation tells us that
$$\ket{c(T)}
 =\exp\!\left(-iBX\int_0^T
       f(t)\cos(\omega(t+\tau)+\phi)\,dt\right)\ket{0} \,.
$$
Since the final operation in our protocol (before the final measurement) is to apply $C_m^{\dagger}$, the final state is $C_m^{\dagger} \ket{\psi(T)} = C_m^{\dagger} C_m \ket{c(T)} = \ket{c(T)}$. So, the outcome distribution of our protocol is that of measuring $\ket{c(T)}$ in the standard basis. By \eqref{eq:averaged-detection-probability}, when $T\geq\sqrt N/(\gamma B)$,
the weight of this state on $\ket1$, averaged over the independent
choices of $f=f_b$ and $\tau$ in Protocol~\ref{prot:main}, is at least
$1/(8\pi)$, as desired. 

The fact that $\ket{c(0)} = \ket{0}$ (before and after the pulse $C_0$) simply follows from the fact that there has been no signal Hamiltonian evolution yet. We are left with showing that $\ket{c(t)}$ remains
continuous across pulses, and satisfies \eqref{eq:coeff-eq}.

Since $\ket{\psi(t)}=C_k\ket{c(t)}$ on interval $(t_{k-1}, t_k)$, and $C_k$ is constant, the
Schrodinger equation for $\ket{\psi(t)}$ gives
\begin{equation}
\label{eq:schrodinger-physical}
 iC_k\frac{d}{dt}\ket{c(t)}
 = H_{sig}(t) C_k \ket{c(t)} = B\cos(\omega(t+\tau)+\phi)ZC_k\ket{c(t)}\,,
\end{equation}
which we can rewrite as
\[
 i\frac{d}{dt}\ket{c(t)}
=B\cos(\omega(t+\tau)+\phi)C_k^\dagger ZC_k\ket{c(t)}.
\]
Thus, for the target equation~\eqref{eq:coeff-eq} to hold on the interval, it suffices for $C_k$ to satisfy
\begin{equation}
 C_k^\dagger ZC_k=\sigma_kX.
 \label{eq:pulse-frame-condition}
\end{equation}
For our protocol, $C_k = G_{k-1} \ldots G_0$ where $G_j = Y$ for $j\geq 1$ and $G_0 = e^{i \sigma_1 \pi Y/4}$. So, $C_k = Y^{k-1} e^{i \sigma_1 \pi Y/4}$. One can verify directly that 
$$ C_1^{\dagger} Z C_1 = e^{-i \sigma_1 \pi Y/4} Z e^{i \sigma_1 \pi Y/4} = \sigma_1 X $$
(it helps to use the identity $e^{i\theta Y} = \cos(\theta)Y + i\sin(\theta)Y$). The identity for larger $k$ follows by induction. Suppose $C_k^{\dagger} Z C_k = \sigma_k X$. Then, $$C_{k+1}^{\dagger} Z C_{k+1} = C_k^{\dagger} Y Z Y C_k = C_k^{\dagger} (-Z) C_k  = -\sigma_k X = \sigma_{k+1} X \,,$$
where the last equality uses the fact that the sign flips at each interval. 

Finally, at a boundary between consecutive intervals, the physical state changes according to
$\ket{\psi(t_k^+)}=G_k\ket{\psi(t_k^-)}$, where $\psi(t_k^-)$ and $\psi(t_k^+)$ denote the physical states right before and right after $G_k$ is applied. So,
\[
 \ket{c(t_k^+)}
 =C_{k+1}^\dagger G_k C_k\ket{c(t_k^-)}
 =C_{k+1}^\dagger C_{k+1}\ket{c(t_k^-)} = \ket{c(t_k^-)}.
\]
Therefore, the coefficient vector is continuous across each boundary.

\section{A discrete Grover-Heisenberg limit}
\label{sec:discrete-lower-bound}

We now prove that the $\Theta(\sqrt m/B)$ sensing time is optimal
in the worst case when the set of possible frequencies is of size $m$. In particular, the explicit family $\omega_j=jB$, $1\leq j\leq m$, gives a matching lower bound
for our upper bound. We refer to this lower bound as the \emph{discrete} Grover-Heisenberg limit.

The proof of this lower bound follows the strategy in \cite{allen2025quantum}, which is to first bound the change of the state on a short time interval, on average over the unknown frequency. Then, this is extended to the full time evolution. We adapt the argument of~\cite{allen2025quantum} by averaging over
the finite set of candidate frequencies rather than a continuous
interval. and we bound the resulting sum using an elementary inequality
for Fourier integrals at separated frequencies.

\begin{theorem}[discrete-frequency sensing lower bound]
\label{thm:discrete-sensing-lower-bound}
Let $B>0$ and let $\omega_1,\ldots,\omega_m>0$ satisfy
\[
 |\omega_i-\omega_j|\geq B\qquad(i\neq j).
\]
Consider a protocol that distinguishes the zero Hamiltonian from each of the signal Hamiltonians
\[
 H_{\mathrm{sig}}(t)=B\cos(\omega_i t+\phi)Z,
 \qquad i\in\{1,\ldots,m\},
\]
with error probability at most $1/3$ on every such Hamiltonian.
Suppose the protocol, including all waits and repetitions,
finishes within time $\mathcal T$.
Then,
\begin{equation}
 \mathcal T\geq\frac{1}{6\sqrt{30}\,e}\frac{\sqrt m}{B}.
 \label{eq:discrete-time-lower-bound}
\end{equation}
The bound holds even if the phase is known to be $\phi=0$, and even
with arbitrarily many auxiliary qubits.
\end{theorem}

We will need the following technical lemma about Fourier integrals at separated frequencies.

\begin{lemma}[Fourier integrals at separated frequencies]
\label{lem:separated-fourier-bound}
Let $I$ be an interval of length $h>0$, and let real numbers
$\lambda_1,\ldots,\lambda_m$ have pairwise separation at least $1/h$.
For any square-integrable function $g:I\to\cH$ taking values in
a Hilbert space,
\begin{equation}
 \sum_{j=1}^m
 \left\|\int_I e^{i\lambda_j s}g(s)\,ds\right\|^2
 \leq 30h\int_I\|g(s)\|^2\,ds.
 \label{eq:separated-fourier-bound}
\end{equation}
The same bound holds with $e^{i\lambda_j s}$ replaced by
$\cos(\lambda_j s)$.
\end{lemma}
\begin{proof}
Fix $g$ and set
\[
 v_j:=\int_I e^{i\lambda_j s}g(s)\,ds,
 \qquad S:=\sum_{j=1}^m\|v_j\|^2.
\]
The quantity $S$ is exactly the left-hand side of
\eqref{eq:separated-fourier-bound}. By the definition of $v_j$,
\begin{align*}
 S
 &=\left|\sum_{j=1}^m
       \left\langle v_j,\int_I e^{i\lambda_j s}g(s)\,ds
       \right\rangle\right|\\
 &=\left|\int_I
       \left\langle\sum_{j=1}^m e^{-i\lambda_j s}v_j,g(s)
       \right\rangle ds\right|.
\end{align*}
Cauchy--Schwarz therefore gives
\begin{equation}
 S^2\leq
 \left(\int_I\|g(s)\|^2\,ds\right)
 \left(\int_I\left\|\sum_{j=1}^m
                 e^{-i\lambda_j s}v_j\right\|^2ds\right).
 \label{eq:fourier-cauchy-schwarz}
\end{equation}
From here, we focus on bounding the second factor.

Let $c$ be the midpoint of $I$ and set
\[
 w(s):=2\max\{1-|s-c|/h,0\}.
\]
This nonnegative function is at least $1$ on $I$. To bound its
Fourier transform, let $J=[-h/2,h/2]$ and let $1_J$ denote its indicator
function. Then
\[
 w(s)=\frac2h\int_J1_J(s-c-u)\,du.
\]
Indeed, the integrand equals $1$ precisely when
\[
 s-c-u\in J
 \quad\Longleftrightarrow\quad
 u\in[s-c-h/2,s-c+h/2].
\]
Since we integrate over $u\in J$, the integral is the length
of the intersection
\[
 [-h/2,h/2]\cap[s-c-h/2,s-c+h/2].
\]
These intervals both have length $h$, and their centers are
distance $|s-c|$ apart. Their overlap therefore has length
$\max\{h-|s-c|,0\}$; multiplying by $2/h$ gives $w(s)$. 

Changing variables to $v=s-c-u$ and interchanging
the integrals gives
\begin{align*}
 \widehat w(\xi)
 :=\int_{\R}w(s)e^{-i\xi s}\,ds
 &=\frac2h\int_J\int_J e^{-i\xi(c+u+v)}\,dv\,du\\
 &=\frac2h e^{-i\xi c}
   \left(\int_J e^{-i\xi u}\,du\right)^2.
\end{align*}
At $\xi=0$, the integral over $J$ is $h$, so $\widehat w(0)=2h$.
For $\xi\neq0$, we have
\[
 \left|\int_J e^{-i\xi u}\,du\right|
 =\left|\frac{e^{-i\xi h/2}-e^{i\xi h/2}}{-i\xi}\right|
 \leq\frac2{|\xi|} \,.
\]
Consequently, $|\widehat w(\xi)|\leq(2/h)(2/|\xi|)^2
=8/(h\xi^2)$. 

Now, order the frequencies $\lambda_i$ in increasing order. Then
$|\lambda_i-\lambda_j|\geq |i-j|/h$, and we have that, for all $i$,
\begin{equation}
\label{eq:30}
|\widehat w(\lambda_i-\lambda_j)| \leq 2h+16h\sum_{r=1}^{\infty}\frac1{r^2}
 =\left(2+\frac{8\pi^2}{3}\right)h
 <30h \,,
\end{equation}
where we used the classic identity
$\sum_{r=1}^{\infty}r^{-2}=\pi^2/6$.

We can now bound the second factor in~\eqref{eq:fourier-cauchy-schwarz}:
\begin{align}
 \int_I\left\|\sum_{j=1}^m e^{-i\lambda_j s}v_j\right\|^2ds
 &\leq\int_{\R}w(s)
       \left\|\sum_{j=1}^m e^{-i\lambda_j s}v_j\right\|^2ds\nonumber\\
 & =\sum_{i,j} \int_{\R} w(s) \,e^{-i(\lambda_i-\lambda_j)s} 
   \langle v_j,v_i\rangle  \, ds\nonumber \\
 &= \sum_{i,j}\widehat w(\lambda_i-\lambda_j)
       \langle v_j,v_i\rangle \nonumber\\
 &\leq\sum_{i,j}
       |\widehat w(\lambda_i-\lambda_j)|\,\|v_i\|\,\|v_j\|\nonumber\\
 &\leq30h\sum_{j=1}^m\|v_j\|^2=30hS.
 \label{eq:separated-exponential-sum}
\end{align}
The first inequality uses $w\geq1$ on $I$ and $w\geq0$ everywhere; the first equality comes from expanding the squared norm; the second inequality is Cauchy-Schwarz; and the last inequality uses
$2\|v_i\|\|v_j\|\leq\|v_i\|^2+\|v_j\|^2$
and the row-sum bound.
The equality includes an absolute value because the preceding
integral is nonnegative. The next two inequalities are the triangle
inequality and Cauchy--Schwarz, respectively. For the last inequality, use
$2\|v_i\|\|v_j\|\leq\|v_i\|^2+\|v_j\|^2$
and the previous bound \eqref{eq:30}.

Substituting~\eqref{eq:separated-exponential-sum}
into~\eqref{eq:fourier-cauchy-schwarz} gives
\[
 S^2\leq30hS\int_I\|g(s)\|^2\,ds.
\]
If $S=0$, the claim is immediate. Otherwise, dividing by $S$ gives the desired bound~\eqref{eq:separated-fourier-bound}.

Finally, for the cosine version, let
$v_j^{\pm}:=\int_I e^{\pm i\lambda_j s}g(s)\,ds$.
Since $\cos(\lambda_j s)=(e^{i\lambda_j s}+e^{-i\lambda_j s})/2$,
\begin{align*}
 \sum_j\left\|\int_I\cos(\lambda_j s)g(s)\,ds\right\|^2
 &=\sum_j\left\|\frac{v_j^++v_j^-}{2}\right\|^2\\
 &\leq\frac12\sum_j\bigl(\|v_j^+\|^2+\|v_j^-\|^2\bigr)\\
 &\leq30h\int_I\|g(s)\|^2\,ds.
\end{align*}
The last step applies the exponential bound separately to the lists
$(\lambda_j)_j$ and $(-\lambda_j)_j$, each of which has the required
separation. 
\end{proof}

With Lemma~\ref{lem:separated-fourier-bound} in hand, we can bound how much an initial state can change over a short time interval, on average over any set of frequencies that are pairwise separated by $B$.
Note that we do not restrict to a single qubit here: our bound holds for any number number of auxiliary qubits, with arbitrary control unitaries on all qubits, and is independent of the dimension.
\begin{lemma}[Bounding how much a state can change on a short interval]
\label{lem:discrete-short-interval}
Let $\cH$ be any finite-dimensional Hilbert space.
Assume $|\omega_i-\omega_j|\geq B$ for $i\neq j$.
Let $I=[a,a+\ell]$, where $0<\ell\leq1/B$, and let
$(M(s))_{s\in I}$ be a measurable family of Hermitian operators on
$\cH$ with $\|M(s)\|\leq1$ for every $s\in I$.
Write $V_j^I$ for the unitary evolution on $\cH$ generated over time interval $I$ by the Hamiltonian
$B\cos(\omega_j s)M(s)$. For every unit vector $v\in\cH$,
\begin{equation}
 \sum_{j=1}^m\|(V_j^I-I)v\|^2
 \leq30B\ell\,e^{2B\ell}.
 \label{eq:discrete-short-interval}
\end{equation}
\end{lemma}
\begin{proof}
The Dyson expansion gives $V_j^I-I=\sum_{r=1}^{\infty}D_{r,j}$,
where
\[
 D_{r,j}v=(-iB)^r
 \int_{a<s_r<\cdots<s_1<a+\ell}
 \left(\prod_{\nu=1}^r\cos(\omega_j s_\nu)\right)
 M(s_1)\cdots M(s_r)v\,ds_r\cdots ds_1.
\]
It converges absolutely since $\|D_{r,j}\|\leq(B\ell)^r/r!$.
To bound all frequencies together, equip the space $\cH^m$ of
$m$-tuples of vectors with the norm
\[
 \|(z_1,\ldots,z_m)\|
 :=\left(\sum_{j=1}^m\|z_j\|^2\right)^{1/2}.
\]
First fix the outer variables $s_1,\ldots,s_{r-1}$ and perform
the innermost integration. Its vector is
\[
 \int_a^{s_{r-1}}\cos(\omega_j u)M(u)v\,du.
\]
For $r=1$, the upper endpoint is $a+\ell$ and there are no outer
variables. Extend the function being integrated by zero to the
interval $[a,a+1/B]$. Its squared $L^2$ norm is at most $\ell$,
because $\|M(u)v\|\leq1$. Lemma~\ref{lem:separated-fourier-bound},
with $h=1/B$, therefore bounds the sum of the squared norms of
these inner integrals by $30\ell/B$.

For $r\geq2$, write $\vec{s}=(s_1,\ldots,s_{r-1})$ and define
the outer integration region
\[
 \Delta_{r-1}
 :=\{a<s_{r-1}<\cdots<s_1<a+\ell\}.
\]
For $\vec{s}\in\Delta_{r-1}$, let
\[
 z_j(\vec{s})
 :=\left(\prod_{\nu<r}\cos(\omega_j s_\nu)\right)
 M(s_1)\cdots M(s_{r-1})
 \int_a^{s_{r-1}}\cos(\omega_j u)M(u)v\,du.
\]
Each cosine has absolute value at most $1$, and
$\|M(s_1)\cdots M(s_{r-1})\|\leq1$. Multiplying the inner
integral by these factors therefore cannot increase its norm.
The preceding bound gives, for every $\vec{s}\in\Delta_{r-1}$,
\[
 \|(z_1(\vec{s}),\ldots,z_m(\vec{s}))\|^2
 =\sum_{j=1}^m\|z_j(\vec{s})\|^2
 \leq\sum_{j=1}^m
   \left\|\int_a^{s_{r-1}}\cos(\omega_j u)M(u)v\,du\right\|^2
 \leq\frac{30\ell}{B}.
\]
Since
$D_{r,j}v=(-iB)^r\int_{\Delta_{r-1}}z_j(\vec{s})\,d\vec{s}$,
the triangle inequality for integration in $\cH^m$ (which is taken componentwise) gives
\begin{align}
 \left(\sum_{j=1}^m\|D_{r,j}v\|^2\right)^{1/2}
 &=B^r\left\|\int_{\Delta_{r-1}}
       (z_1(\vec{s}),\ldots,z_m(\vec{s}))\,d\vec{s}\right\| \nonumber\\
 &\leq B^r\int_{\Delta_{r-1}}
       \|(z_1(\vec{s}),\ldots,z_m(\vec{s}))\|\,d\vec{s} \nonumber\\
 &\leq B^r\sqrt{\frac{30\ell}{B}}\,
       \operatorname{vol}(\Delta_{r-1}) \nonumber\\
 &=B^r\sqrt{\frac{30\ell}{B}}\,
       \frac{\ell^{r-1}}{(r-1)!}. \label{eq:dyson-bound}
\end{align}
For $r=1$, the same estimate follows directly from the
inner-integral bound, with no outer integration and a volume
factor of $1$.
Finally, the Dyson expansion gives
$(V_j^I-I)v=\sum_{r\geq1}D_{r,j}v$ for each $j$.
Collecting all frequencies into one tuple, we therefore have
\[
 \bigl((V_1^I-I)v,\ldots,(V_m^I-I)v\bigr)
 =\sum_{r=1}^{\infty}
   \bigl(D_{r,1}v,\ldots,D_{r,m}v\bigr).
\]
Apply the triangle inequality to this infinite sum (which is fine since the Dyson series converges absolutely), and then use our bound \eqref{eq:dyson-bound}:
\begin{align*}
 \left(\sum_{j=1}^m\|(V_j^I-I)v\|^2\right)^{1/2}
 &\leq\sum_{r=1}^{\infty}
       \left(\sum_{j=1}^m\|D_{r,j}v\|^2\right)^{1/2}\\
 &\leq\sum_{r=1}^{\infty}
       B^r\sqrt{\frac{30\ell}{B}}\,
       \frac{\ell^{r-1}}{(r-1)!}\\
 &=\sqrt{30B\ell}
       \sum_{r=1}^{\infty}\frac{(B\ell)^{r-1}}{(r-1)!}\\
 &=\sqrt{30B\ell}\,e^{B\ell}.
\end{align*}
Squaring proves~\eqref{eq:discrete-short-interval}. Notice that we did not
assume any commutativity of the operators $M(s)$.
\end{proof}

\begin{proof}[Proof of Theorem~\ref{thm:discrete-sensing-lower-bound}]
It suffices to consider the easier problem in which the phase is
known to be $\phi=0$. Our lower bound holds even when the phase is known. 

We can purify the entire protocol, and represent it as a unitary evolution on a
larger Hilbert space followed by one final measurement. This preserves the protocol's output distribution.

As in Section~\ref{sec:analysis_one}, let $U_c(t)$ denote the
control-only evolution, now on the enlarged Hilbert space.
Here $t$ is absolute time from the start of the protocol,
including all waits. Following~\eqref{eq:1}, define the coefficient
vector in the control-following basis by $\ket{c(t)} =U_c(t)^\dagger\ket{\psi(t)}$. 
For ease of notation, denote $v:=\ket{c(0)}$. The very same calculation
leading to~\eqref{eq:ct} applies on this enlarged space with
$Z\otimes I$ in place of $Z$. If the signal is absent, we have
$\ket{c(t)}=v$ for all $t$. For a signal of frequency $\omega_j$,
write $\ket{c_j(t)}$ for the coefficient vector and $V_j(t)$ for
its unitary evolution, so that $\ket{c_j(t)}=V_j(t)v$.
Since we have fixed the phase $\phi$ to be zero, these unitaries satisfy
\begin{equation}
 i\frac{d}{dt}V_j(t)
 =B\cos(\omega_j t)M(t)V_j(t),
 \qquad
 M(t)=U_c(t)^\dagger(Z\otimes I)U_c(t),
 \qquad V_j(0)=I.
 \label{eq:lower-bound-interaction-picture}
\end{equation}

\vspace{2mm}
By hypothesis, the protocol we are considering has a probability of error at most $1/3$. Thus, the two output distributions when the signal is absent and when it is present have total variation
distance at least $1/3$. For normalized pure
states, the trace distance is at most their vector distance. 
\[
 \frac12\bigl\|\ket{x}\bra{x}-\ket{y}\bra{y}\bigr\|_1
 =\sqrt{1-|\langle x|y\rangle|^2}
 \leq\|x-y\|.
\]
Let $\mathcal T$ denote the final time. Then, since measurement and tracing out cannot increase trace distance, we must have
\begin{equation}
 \|V_j(\mathcal T)v-v\|\geq\frac13
 \quad(1\leq j\leq m),
\end{equation}
The latter is because $V_j(\mathcal T)v$ and $v$ are respectively the final states of the protocol when the signal is present and when it is absent, expressed in the control-following basis.
Then, we have 
\begin{equation}\mathcal D(\mathcal T):=
 \frac1m\sum_{j=1}^m\|V_j(\mathcal T)v-v\|^2
 \geq\frac19
 \label{eq:lower-bound-required-distance}
\end{equation}

We will now show a competing upper bound on $\mathcal D(\mathcal T)$ due to Lemma~\ref{lem:discrete-short-interval}, which eventually leads to the desired upper bound on $\mathcal T$.

First, note that integrating
\eqref{eq:lower-bound-interaction-picture} and using unitarity gives
\[
 \|V_j(\mathcal T)v-v\|
 \leq B\int_0^{\mathcal T}
       |\cos(\omega_j t)|\,\|M(t)V_j(t)v\|\,dt
 \leq B\mathcal T.
\]
Equation~\eqref{eq:lower-bound-required-distance} therefore implies
$B\mathcal T\geq1/3$.

Now, divide $[0,\mathcal T]$ into
$q=\lceil B\mathcal T\rceil$ equal intervals of length
$\ell=\mathcal T/q\leq1/B$. Let $V_{j,k}$ be the unitary evolution
on interval $k$, so $V_j(\mathcal T)=V_{j,q}\cdots V_{j,1}$.
Note that we have the telescoping identity
\[
 V_{j,q}\cdots V_{j,1}-I
 =\sum_{k=1}^q
 V_{j,q}\cdots V_{j,k+1}(V_{j,k}-I)
\]
where the empty product when $k=q$ is equal to $I$.
After applying the above to $v$, unitarity of each $V_{j,q}\cdots V_{j,k+1}$ and a triangle inequality yield
\begin{align*}
 \left(\sum_{j=1}^m\|V_j(\mathcal T)v-v\|^2\right)^{1/2}
 &\leq\sum_{k=1}^q
       \left(\sum_{j=1}^m\|(V_{j,k}-I)v\|^2\right)^{1/2}\\
 &\leq q\sqrt{30B\ell}\,e^{B\ell}\,,
\end{align*}
where the last inequality is by Lemma~\ref{lem:discrete-short-interval}.
Crucially, every difference $V_{j,k}-I$ acts on the same vector
$v$, rather than on a frequency-dependent intermediate state.
Squaring and dividing by $m$ gives
\begin{equation}
 \mathcal D(\mathcal T) \leq \frac{30q^2 B \ell e^{2B\ell}}{m}
 \leq\frac{30qB\mathcal T}{m}e^{2B\mathcal T/q}
 \leq\frac{30e^2}{m}B\mathcal T(B\mathcal T+1),
 \label{eq:lower-bound-distance-growth}
\end{equation}
where we used $B\mathcal T/q\leq1$ and $q\leq B\mathcal T+1$.

Since $B\mathcal T\geq1/3$, we have
$B\mathcal T+1\leq4B\mathcal T$. Combining the two bounds
on $\mathcal D$ now gives
\[
 \frac19\leq\frac{120e^2B^2\mathcal T^2}{m}.
\]
Rearranging proves~\eqref{eq:discrete-time-lower-bound}.
\end{proof}

\begin{remark}
We remark that the $\Omega(\sqrt m/B)$ lower bound is only a worst-case statement over sets of $m$ candidate frequencies, and cardinality alone of course does not imply this lower bound for every set. For example, the candidate frequencies could be clustered in a very small interval, say of size $O(B)$. Then, even a conventional single-qubit
protocol for a continuous window can solve the sensing problem in $O(1/B)$ independently of $m$.
\end{remark}

\bibliographystyle{alpha}
\bibliography{references}

\appendix
\section{Proof of Lemma~\ref{lem:packing-maximizer-exists}}
\label{sec:app}
We prove Lemma~\ref{lem:packing-maximizer-exists}, which establishes the existence of a maximizer for the functional $J$ considered in Section~\ref{sec:analysis_packing}. We restate the lemma here for convenience.

\begin{maximizer-restated}[Existence of a maximizer]
For every $C>0$, the functional $J$ attains a finite maximum
on $\bT$.
\end{maximizer-restated}
\begin{proof}
Let $\mathcal K = \{v\in L^2([0,2\pi];\mathbb C):
                 |v(x)|\leq1\text{ almost everywhere}\}$
be the closure of $\mathbb T$ in $L^2$ norm. Define $J$ on $\mathcal K$ by the right-hand side of
Equation~\eqref{eq:J}, with the convention $\log0=-\infty$. We say a sequence of functions $(f_n)$ in $\mathcal K$ weakly
converges to $f \in \mathcal K$ if for any
$h \in L^2([0,2\pi];\mathbb C)$, we have
$\braket{h,f_n}\to\braket{h,f}$.
We need the following property of the functional $J$:

\begin{lemma}\label{lem:J-weakly-upper-semicontinous}
    Suppose $(f_n)$ weakly converges to $f$. Then, we have
    \begin{align*}
        J(f) \geq \limsup_{n\to\infty} J(f_n).
    \end{align*}
    (For the reader familiar with functional analysis, this lemma
    says that the functional $J$ is weakly upper semicontinuous).
\end{lemma}
\begin{proof}[Proof of Lemma~\ref{lem:J-weakly-upper-semicontinous}]
    If there exists an $i$ satisfying $\iprod{g_i, f} = 0$, then $J(f) = -\infty$, but also, since all the terms in $J$ are non-positive,
    \begin{align*}
        J(f_n)
        \leq \log \left| \iprod{g_i, f_n} \right|
        \to -\infty,
    \end{align*}
    so the statement of the lemma holds with equality.

    Therefore we can assume that for all $i$ the correlation
    $\braket{g_i,f}$ is nonzero. Then,
    $\lim_{n\to\infty} \log\left|\iprod{g_i, f_n}\right|
    = \log\left|\iprod{g_i,f}\right|$
    is a finite value for all $i$, so it is enough to focus on the penalty term and prove that
    \begin{align*}
        P(f) \geq \limsup_n P(f_n),
    \end{align*}
    where
    $P(f) = \frac{1}{2\pi}\int_0^{2\pi}
    \log (1-|f(x)|^2) dx$.

    To prove $P(f) \geq \limsup_n P(f_n)$,
    we will first prove that
    $P(rf) \geq \limsup_n P(f_n)$
    for all $0<r<1$, and
    conclude the argument with
    $\lim_{r\uparrow1}P(rf)=P(f)$.
    Let
    $h_r(x):=\frac{f(x)}{1-r^2|f(x)|^2}$. 
    Since the denominator is at least $1-r^2>0$ almost
    everywhere, $h_r\in L^2$.
    Then we have
    \begin{align*}
        P(f_n) - P(rf)
        &= \frac{1}{2\pi} \int_{0}^{2\pi}
        \log \frac{1-|f_n(x)|^2}{1-r^2|f(x)|^2} dx\\
        &\leq \frac{1}{2\pi} \int_{0}^{2\pi}
        \left(\frac{1-|f_n(x)|^2}{1-r^2|f(x)|^2}-1\right) dx\\
        &= \frac{1}{2\pi} \int_{0}^{2\pi}
        \frac{r^2|f(x)|^2-|f_n(x)|^2}
             {1-r^2|f(x)|^2} dx\\
        &= -\frac{1}{2\pi}\int_0^{2\pi}
        \frac{|f_n(x)-rf(x)|^2
        +2r\operatorname{Re}\!\left(
            \overline{f(x)}(f_n(x)-rf(x))
        \right)}
        {1-r^2|f(x)|^2}\,dx\\
        &\leq -2r\operatorname{Re}
        \langle h_r,f_n-rf\rangle\\
        &\leq -2r\operatorname{Re}
        \langle h_r,f_n-f\rangle,
    \end{align*}
    where the inequality in the second line is the basic
    inequality $\log t \leq t-1$, and the inequality in the
    last line is because
    $
    \iprod{h_r,f}
    =\frac1{2\pi}\int_0^{2\pi}
      \frac{|f(x)|^2}{1-r^2|f(x)|^2}\,dx
    \geq 0$.

    Now we can fix $r$ and let $n\to\infty$. According to our
    assumption $\iprod{h_r, f_n-f} \to 0$. Therefore
    $P(rf)\geq \limsup_n P(f_n)$.
    Finally, letting $r\uparrow1$ proves the lemma.
\end{proof}

Now we can prove Lemma~\ref{lem:packing-maximizer-exists}.
Let $(f_n)$ be a sequence of functions in $\mathcal K$ that
approaches the supremum of $J$.
The set $\mathcal K$ is a bounded, closed and a convex subset
of the Hilbert space $L^2$, so it is weakly sequentially compact.
This means that there exists a subsequence $(f_{n_j})$ that weakly converges
to some $f \in \mathcal K$, therefore according to
Lemma~\ref{lem:J-weakly-upper-semicontinous},
$J(f) \geq \limsup_j J(f_{n_j}) = \sup J$.
So $f$ is exactly a maximizer of $J$.

The maximizer $f$ has a representative in
$\mathbb T$.
If $|f(x)|\geq1$ only on a set of measure zero, we can just set the function to zero at these points to get a maximizer in $\mathbb T$.
If $|f(x)|=1$ for a positive measure set,
then $J(f)=-\infty$, but this is impossible as
the supremum of $J$ is finite.\footnote{
To see this, take $f_0 = \frac12 \mathbf1_{[0,\eps]}$ for some small
enough $\eps$. Then, all the $g_i$ are positive over
$[0,\eps]$, so all correlations $\iprod{g_i,f_0}$ are nonzero, 
and so $J(f_0)>-\infty$. Moreover, $J(f)\leq0$ for all $f$, since all the terms in $J$ are non-positive. So, the supremum of $J$ is finite.}
\end{proof}

\section{Proof of Theorem~\ref{thm:algorithmic}}
\label{app:algorithmic}
Here we describe the algorithm that achieves the guarantees in Theorem~\ref{thm:algorithmic}, and prove that it achieves said guarantees.

    Let $u_0(x)$ be the function which is $1/2$ for $x \in [0, \tfrac{1}{8 \pi W}]$ and $0$ elsewhere.
    Then $|\iprod{g_i, u_0}| = \iprod{g_i, u_0} \geq \Omega (\log W)$ for all $i \in [N]$, and hence
    \begin{align*}
        J(u_0) &\geq -\Omega \left( N \log W \right) + C N \frac{1}{2 \pi} \int_0^{1 / (8 \pi W)} \log (3/4) dx \\
        &= -\Omega \left( N \log W \right) \; .
    \end{align*}
        
    Consider the following randomized, iterative procedure.
    Let $u_0$ be as above, and initialize $t = 0$.
    Then, repeatedly perform the following operations:
    \begin{enumerate}
        \item Check if $|\iprod{g_i, u_t}| \geq c / \sqrt{N}$ for all $i \in [N]$.
        If this is true, return $u_t$.
        \item Else, let $i$ be an index which violates the inequality. Let $r$ be as in Corollary~\ref{cor:potential-progress}, let $z$ be uniformly random from $B_r$, and let $v_{t + 1} = \Phi_z^{(i)} (u_t)$.
        \item If $\norm{v_{t + 1}}^2 > 1 / (2C)$, let $\tilde{v}_{t + 1} = \frac{1}{\sqrt{2C}} \cdot \frac{v_{t + 1}}{\norm{v_{t + 1}}}.$
        \item If $J(\tilde{v}_{t + 1}) - J(u_t) \geq 0$, let $u_{t + 1} = \tilde{v}_{t + 1}$.
        Otherwise, let $u_{t + 1} = u_t$.
        \item Set $t \gets t + 1$.
    \end{enumerate}
    Clearly, if this protocol terminates at some iteration $T$, then $u_T$ satisfies the desired inequalities.
    It therefore suffices to show that the expected number of iterations is bounded.
    By our choice of $W$, we have that $\norm{u_0}^2 \leq 1 / (2C)$, and so by construction, $\norm{u_t}^2 \leq 1 / (2C)$ for all $t$.
    Additionally, observe that $J(u_{t + 1}) \geq J(u_t)$ deterministically, for all $t$.
    Moreover, by Lemma~\ref{lem:packing-energy-bound} and Corollary~\ref{cor:potential-progress}, we have that, as long as the algorithm has not terminated, 
    \[
    \mathbb{E} [J(u_{t + 1}) - J(u_t) | u_t] \geq \mathbb{E} [J(v_{t + 1}) - J(u_t) | u_t] \geq \frac{1}{2} \log 2 \; ,
    \]
    and hence, if we treat $T$ as a stopping time, we have that 
    \[
    \Omega \left( N \log (1 / W) \right) \geq \mathbb{E} [J(u_{T}) - J(u_0)] \geq \frac{1}{2} \log 2 \cdot \mathbb{E} [T] \; ,
    \]
    so the expected number of total iterations is at most $\Omega \left( N \log (1 / W) \right)$.

    We now turn to the claimed structure of the output.
    Let $z_1 \leq z_2 \leq \ldots z_M$ denote all zeroes of all $g_i$ for all $i \in [N]$ in the range $[0, 1]$ as well as all integer multiples of $W$ in the same range, let $z_0 = 0$ and $z_{M + 1} = 1$, and let $\mathcal{I} = \{ [z_i, z_{i + 1}) \}_{i = 0}^{m}$ denote the partition of the identity into disjoint intervals with these endpoints.
    Note that there are at most $O(N W)$ total such intervals, and this partition can be computed in time $O(N W \log (NW)) \ll O(N^3 W \log W)$, by sorting the zeroes of all the cosines. 
    Observe additionally that $u_0$ is piecewise constant across these intervals.
    Moreover, if $u$ is any function which is piecewise constant on these intervals, then so is $\Phi^{(i)}_z (u)$, since on any interval, the sign of $\cos (\omega_i x)$ cannot change, and thus $L_{s_i (x), u(x)} (z)$ is constant.
    
    Finally, we prove the desired bound on the runtime of this method. 
    Observe that for any function $f$ which is piecewise constant on $\mathcal{I}$, computing $\Phi_z^{i} (f)$ can be done in $O(N W)$ time, and evaluating $\iprod{g_i, f}$ can be done in $O(N W)$ time for any $i$.
    Thus, each iteration of the loop can be performed in time $O(N^2 W)$, and combining this with the bound on the expected number of iterations yields the total expected runtime of $O(N^3 W \log (W))$, as claimed.

\end{document}